\documentstyle[10pt,a4wide]{article}
\newcommand{\ol}{\overline}			
\newcommand{\wt}{\widetilde}			
\newcommand{\bdm}{\begin{displaymath}}
\newcommand{\edm}{\end{displaymath}}
\newcommand{\beq}{\begin{equation}}
\newcommand{\eeq}{\end{equation}}
\newcommand{\bea}{\begin{eqnarray}}
\newcommand{\eea}{\end{eqnarray}}
\newcommand{\beao}{\begin{eqnarray*}}
\newcommand{\eeao}{\end{eqnarray*}}
\newcommand{\de}{{\rm d}}
\newcommand{\e}{\varepsilon}
\newcommand{\f}{\phi}
\newcommand{\g}{\gamma}
\newcommand{\r}{\varrho}
\newcommand{\n}{\nu}
\newcommand{\tot}[2]{\frac{{\rm d} #1}{{\rm d} #2}}		
\newcommand{\prt}[2]{\frac{\partial #1}
                     {\partial #2}}  		
\def\abs#1{\left\vert #1 \right\vert}		
\def\rund#1{\left( #1 \right)}                  
\def\RUND#1{\Bigl( #1 \Bigr)}			
\def\eck#1{\left\lbrack #1 \right\rbrack}	
\def\eckk#1{\bigl[ #1 \bigr]}			
\def\ECK#1{\Bigl[ #1 \Bigr]}			
\def\wave#1{\left\lbrace #1 \right\rbrace}	
\def\ave#1{\left\langle #1 \right\rangle}	
\def\at#1{\left. #1 \right\vert}		
\def\bbbr{{\rm I\!R}}                           
\def\SFB{{This work was supported by the ``Sonderforschungsbereich
375-95 f\"ur
Astro--Teil\-chen\-phy\-sik" der Deutschen For\-schungs\-ge\-mein\-schaft.}}
\catcode`\@=11

\def\@biblabel#1{}
\def\@cite#1#2{#1\if@tempswa , #2\fi}
\def\@citex[#1]#2{\if@filesw\immediate\write\@auxout{\string\citation{#2}}\fi
  \def\@citea{}\@cite{\@for\@citeb:=#2\do
    {\@citea\def\@citea{,\penalty\@m\ }\@ifundefined
       {b@\@citeb}{{\bf ?}\@warning
       {Citation `\@citeb' on page \thepage \space undefined}}%
{\csname b@\@citeb\endcsname}}}{#1}}
\def\thebibliography#1{\section*{\refname\@mkboth
 {\uppercase{\refname}}{\uppercase{\refname}}}\list
 {[\arabic{enumi}]}{\labelsep=1em
 \settowidth\labelwidth{#1}\leftmargin\labelwidth
 \advance\leftmargin\labelsep\itemindent=-\leftmargin
 \itemsep=0.6ex\parsep=0pt\usecounter{enumi}}
 \def\newblock{\hskip .11em plus .33em minus .07em}
 \sloppy\clubpenalty4000\widowpenalty4000
 \sfcode`\.=1000\relax}

\def\@lbibitem[#1]#2{\item[\@biblabel{#1}]\if@filesw 
      { \def\protect##1{\string ##1\space}\immediate
        \write\@auxout{\string\bibcite{#2}{#1}}}\fi\ignorespaces}

\def\@bibitem#1#2{\item\if@filesw \immediate\write\@auxout
       {\string\bibcite{#2}{\the\c@enumi}{#1}}\fi\ignorespaces}

\def\bibcite#1#2{\global\@namedef{b@#1}{#2}}

\catcode`\@=12%
\input psfig.tex
\begin{document}
%
%
%
\title{Reanalysis of the association of high-redshift 1-Jansky quasars
       with IRAS galaxies}
\author{A.~Bartsch, P.~Schneider, and M.~Bartelmann\\
	{\small Max-Planck-Institut f\"{u}r Astrophysik,}\\
	{\small Karl-Schwarzschild-Str.~1,
           D-85740 Garching, Germany}}
\date{January 1996}
\maketitle
%
%
%
\begin{abstract}
We develop a new statistical method to reanalyse angular
correlations between background QSOs and foreground galaxies that are
supposed to be a consequence of dark matter inhomogeneities acting as
weak gravitational lenses. The method is based on a weighted average
over the galaxy positions and is optimized to distinguish between a random
distribution of galaxies 
around QSOs and a distribution which follows an assumed QSO-galaxy
two-point correlation function, by choosing an appropriate weight
function. With simulations we demonstrate that this weighted average
is  slightly more significant than Spearman's rank-order test
which was used in previous investigations. In particular, the
advantages of the weighted average show up if the two-point
correlation function is weak.

We then reanalyze the correlation between high-redshift 1-Jansky QSOs
and IRAS galaxies, taken from the IRAS Faint Source Catalog; these
samples were analyzed previously using Spearman's rank-order test. In
agreement with the previous work, we find moderate to strong
correlations between these two samples; considering the angular
two-point correlation function of these samples, we find a typical
scale of order $5'$ from which most of the correlation signal
derives. However, the statistical significance of the correlation
changes with the redshift slices of the QSO sample one
considers. Comparing with simple theoretical estimates of the expected
correlation, we find that the signal we derive is considerably
stronger than expected. On the other hand, recent direct verifications
of the overdensity of matter in the line-of-sight to high-redshift
radio QSOs obtained from the shear field around these sources,
indicates that the observed association can be attributed to a
gravitational lens effect.
%
%
\end{abstract}
%
%
%
\section{Introduction}
It was argued by Bartelmann \&\ Schneider (\cite{msb91}, \cite{msb92},
\cite{msb93a}, \cite{msb93b}) that a statistical association between
foreground galaxies and distant, radio-loud background sources, as
claimed to be observed by Fugmann (\cite{fug88}, \cite{fug90}), can be
caused by 
gravitational lensing effects due to large-scale structures of the
dark-matter distribution in the Universe. Using Spearman's rank-order
test to investigate the association of 1Jy sources with Lick galaxies
and IRAS galaxies, they indeed found correlations at a high level of
significance (Bartelmann \& Schneider \cite{msb93b}, \cite{msb94},
hereafter BS). A
quantification of such correlations could prove to be a unique tool for
directly probing dark-matter inhomogeneities. It is therefore very
important to verify the results of these analyses with independent
methods and, if possible, to improve their statistical significance, or
to obtain 
more detailed information about the association, e.g. the amplitude of
the correlation function or
a characteristic angular scale. Other groups have shown a statistically
significant association between other samples of QSOs and foreground
matter: Rodrigues-Williams \& Hogan (\cite{rod94}), 
Seitz \& Schneider (\cite{sei95})
and Wu \& Han (\cite{wu95}) have shown evidence for an overdensity of
Abell and Zwicky clusters around high-redshift QSOs, 
and Hutchings (\cite{hut95}) 
has studied the distribution of galaxies around
seven QSOs at $z=2.3$, and
found a statistically significant excess around all of them. Whereas
he did not interpret this result as being due to gravitational
lensing, it appears to be a more natural explanation than assuming
that these galaxies are spatially associated with the QSOs, which
would imply an enormous luminosity evolution.

Direct evidence for the presence of lensing matter in the
line-of-sight towards high-redshift QSOs was obtained by Fort et
al.\ (\cite{for95}) who imaged faint galaxies around several
high-redshift 1Jy QSOs. For several of them, they obtained clear
evidence for a coherent 
shear pattern around these QSOs, which can be attributed to local
concentrations of faint galaxies. These concentrations may indicate
the presence of a group or a cluster, but they are so faint optically
that they would not appear in any cluster catalog. What this might
suggest is that there exists a population of clusters with a much
larger mass-to-light ratio than those clusters which are selected
because of their high optical luminosity, i.e., which appear in
optically-selected cluster catalogs. If these findings are confirmed
(e.g., by HST observations), one has found a way to obtain a {\it
mass-selected sample} of clusters and/or groups.   

In the following sections, we introduce a new approach to test for
correlations (Sect.\ \ref{method}), and use numerical simulations
(Sect.\ \ref{numsim}) to demonstrate that it is applicable and can lead
to an improvement of the significance of the results, if compared to
Spearman's rank-order test. Section~\ref{qsoiras} then presents the
results that are
obtained with our method for 1Jy quasars and IRAS galaxies. In
addition, we perform further simulations to compare our findings with
what is expected from theory. Finally we summarize and present our
conclusions in Sect.\ \ref{summary}.  
%
%
%
\section{Method}
\label{method}
In this section we define our new correlation test, based on a
weighted average. Furthermore we show how it can be applied for
measuring quasar-galaxy associations and how to make use of an a
priori guess (e.g. from theory) of the quasar-galaxy correlation
function to find an optimum weight function. 

For readers who lack the patience to follow the arguments below, we
now give a brief summary of the results of this section, which should
enable him or her to directly go to Sect.~\ref{numqgcorr}.

Given a sample of QSOs, and a sample of galaxies around these QSOs,
such that $\phi_i$ is the angular separation of the $i$-th galaxy from
its associated QSO we then define a correlation coefficient $r_g$ by
\bdm
   r_g=\frac{1}{N}\sum_{i=1}^N  g(\f_i)\ ,
\edm
where $g(\phi)$ is a weight function. Given an assumed two-point
correlation function $\xi_{\rm qg}(\phi)$ between QSOs and galaxies,
we show that the optimal choice of the weight function to allow the
distinction between the assumed two-point correlation function and a
random distribution of galaxies relative to the QSOs is given by
\beq\label{eq40}
   g(\phi)=a\,\xi_{\rm qg}(\phi) +b\quad ,
\eeq
with arbitrary ($a\ne 0$) constants $a$, $b$. 
\subsection{Definitions}
\label{defs}
For any realisation $\rund{\vec{x},\vec{y}}$ of a pair
$\rund{\vec{X},\vec{Y}}$ of $n$-dimensional 
random variables $\vec{X}:=\rund{X_1,\ldots,X_n}$ and  
$\vec{Y}:=\rund{Y_1,\ldots,Y_n}$
we define a correlation coefficient by
\beq\label{eq100}
   r\rund{\vec{x},\vec{y}}:=\sum_{i=1}^{n}
   g\rund{x_i}\cdot f\rund{y_i}
\eeq
with arbitrary functions $g$ and $f$. Formally, this can be understood
as an average of $g\rund{x_i}$ weighted with $f\rund{y_i}$ (or
vice versa), but without the usual normalisation of the weights.
In our application below, the vector $\vec{Y}$ denotes the counts of
galaxies in $n$ concentric rings around quasars, and the vector
$\vec{X}$ denotes the radii of these rings.

Given the probability density $p_{\rm o}\rund{\vec{x},\vec{y}}$ of 
$\rund{\vec{X},\vec{Y}}$ and one
single realisation $(\vec{x}',\vec{y}')$ of two random
variables, it is our aim to decide whether the assumption of
$(\vec{x}',\vec{y}')$ being a realisation of 
$\rund{\vec{X},\vec{Y}}$ can be rejected or not. For this purpose we
use $p_{\rm o}\rund{\vec{x},\vec{y}}$ to determine the distribution
of the correlation coefficient for realisations of
$\rund{\vec{X},\vec{Y}}$, i.e.\ we calculate the cumulative
probability 
\bdm
   P_{\rm o}\rund{r\ge R} = \int\!\!\cdots\!\!\int \de^{n}x\,
   \de^{n}y\, p_{\rm o}\rund{\vec{x},\vec{y}}\,
   \Theta\rund{r\rund{\vec{x},\vec{y}} - R}
\edm
for $r$ taking a value greater than or equal to some threshold $R$
for any realisation of $\rund{\vec{X},\vec{Y}}$.  

Now suppose we find a value $R' := r(\vec{x}',\vec{y}')$ of the
correlation coefficient of $(\vec{x}',\vec{y}')$, and let us define
$\e$ as 
\bdm
   \e := P_{\rm o}\rund{r\ge R'}\ \ .
\edm
Premising $(\vec{x}',\vec{y}')$ to be a realisation of
$\rund{\vec{X},\vec{Y}}$, we know that $\e$ is the probability of the 
correlation coefficient to give a result greater than or equal to $R'$.
Hence, if $\e$ happens to be very small (or very large because then
$1-\e=P_{\rm o}(r<R')$ is very small) we might reject our
premise but rather assume $(\vec{x}',\vec{y}')$ to be drawn from
a different pair of random variables $(\vec{X}',\vec{Y}')$. 

As a consequence of this strategy, we would erroneously conclude for a
fraction $\e$ of all realisations of $\rund{\vec{X},\vec{Y}}$ that
they are not drawn from $\rund{\vec{X},\vec{Y}}$, because
their correlation coefficient is greater than or equal to $R'$.
Throughout this paper the value of $\e$ will therefore be called
the `error level'.

Up to now, we did not specify the functions $g$ and $f$. By intuition
one is lead to the idea that they should be adapted to the given
problem. Imagine we want to check whether $(\vec{x}',\vec{y}')$
can be considered a realisation of $\rund{\vec{X},\vec{Y}}$.
Furthermore, we suspect that $(\vec{x}',\vec{y}')$ has been
drawn from different random variables $(\vec{X}',\vec{Y}')$ with
a corresponding probability density
$p_{\rm a}(\vec{X}',\vec{Y}')$, where the subscript `a' stands for the
`alternative hypothesis'. What we want to achieve then is that the
correlation test allows for an optimum distinction between these two
hypotheses. For both of them we can, in principle, derive the
probability density of the correlation coefficient $r$ from the 
equations
\beao
   p_{\rm o}\rund{r} & = &\int\!\cdots\!\int \de^n x\,\de^n y
   \, p_{\rm o}\rund{\vec{x},\vec{y}}\,
   \delta\rund{r\rund{\vec{x},\vec{y}}-r}\ \ ,\\
   p_{\rm a}\rund{r} & = &\int\!\cdots\!\int \de^n x\,\de^n y
   \, p_{\rm a}\rund{\vec{x},\vec{y}}\,
   \delta\rund{r\rund{\vec{x},\vec{y}}-r}\ \ ,
\eeao
where $\delta\rund{r}$ denotes Dirac's delta function. The first
definition one can think of to quantify the distinction between
$p_{\rm o}\rund{r}$ and $p_{\rm a}\rund{r}$ is the mean error
level 
\beq\label{eq140}
   \ave{P_{\rm o}\rund{r\ge R}}:=
   \int P_{\rm o}\rund{r\ge R}\,p_{\rm a}\rund{R}\,\de R\ \ .
\eeq
According to the preceding explanations the mean error level should be
as small (or large) as possible.

\begin{figure}
   \centerline{\psfig{figure=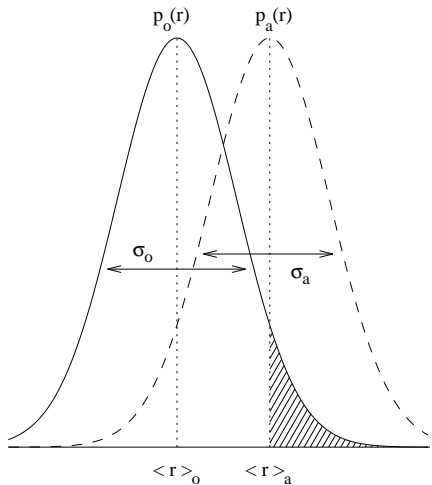,height=7.0cm,width=7.0cm}}
   \caption[]{A simple measure of the distinction of two Gaussian-like 
   distributions is the separation of their mean values in units of
   either one of their standard deviations. The shaded area represents
   the error level $P_{\rm o}\rund{r\ge \ave{r}_{\rm a}}$.}
   \label{fig1}
\end{figure}

If $p_{\rm o}\rund{r}$ and $p_{\rm a}\rund{r}$ are
characterised by a single (e.g. Gaussian-like) peak, we can also
expect the quantities 
\bea
   Q_1 & := & \frac{\ave{r}_{\rm a}-\ave{r}_{\rm o}}
                   {\sigma_{\rm o}}\ \ ,\label{eq150}\\ 
   Q_2 & := & \frac{\ave{r}_{\rm a}-\ave{r}_{\rm o}}
                   {\sigma_{\rm a}}\ \ ,\label{eq160}
\eea
together with the definitions
\beao
   \ave{r}_{\rm a,o} & := & \int r\,p_{\rm a,o}
   \rund{r}\,\de r\ \ ,\\
   \sigma_{\rm a,o} & := & \eck{\int\rund{r-
   \ave{r}_{\rm a,\rm o}}^2 p_{\rm a,o}\rund{r}
   \de r}^{1/2}\ \ ,
\eeao
to be good measures of the distinction between $p_{\rm o}\rund{r}$ and
$p_{\rm a}\rund{r}$. Figure \ref{fig1} illustrates this
concept. In the following section we want to specify the functions $g$
and $f$. We claim that by maximizing either $\abs{Q_1}$ or
$\abs{Q_2}$ we can find $g$ and $f$ such that $r$ as a test for
quasar-galaxy correlations operates close to its optimum sensitivity, 
which we will demonstrate in Sect.\ \ref{numsim} by means of numerical
simulations.  

%
%
%
%
%
\subsection{Quasar-galaxy correlations}
\label{qgcorr}
We now want to apply these theoretical concepts to the associations
between quasars and galaxies by analysing the radial distribution of
galaxies in the vicinity of quasars from a given sample. Suppose that
for each quasar we have a list of relative angular coordinates of
galaxies within a certain (preferably circular) field around the
quasar. 

One way to investigate the galaxy distribution is to merge these lists
into one total list corresponding to a total galaxy field. For the
correlation test we choose an inner and an outer angular radius,
$\f_{\rm in}$ and $\f_{\rm out}$, to define a ring 
$\eck{\f_{\rm in},\f_{\rm out}}$ within the total field which
is divided into $n$ concentric annuli $\eck{\r_i,\r_{i+1}}$ by
the $n+1$ radii
\beq\label{eq180}
   \r_i := \f_{\rm in}+\rund{i-1}\cdot
   \frac{\f_{\rm out} - \f_{\rm in}}{n}\ \ ,\ \ \ i=1,\ldots,n+1\ .
\eeq 
Let $z_i$ denote the number of galaxies within $\eck{\r_i,\r_{i+1}}$
and set $\vec{x}:=\rund{\r_1,\ldots,\r_n}$,
$\vec{y}:=\rund{z_1,\ldots,z_n}$. Then, the correlation coefficient
(\ref{eq100}) reads
\beq\label{eq200}
   r=\sum_{i=1}^n g\rund{\r_i}\,f\rund{z_i}\ \ .
\eeq

If we increase the number $n$ of sub-rings, while the total number
$N:=z_1+\cdots+z_n$ of galaxies within 
$\eck{\f_{\rm in},\f_{\rm out}}$ is constant, we will eventually
reach the situation where each of the sub-rings contains one galaxy at
most, i.e. $z_i=0$ or $1$ for $i=1,\ldots,n$. At that point
$f\rund{z_i}$ in Eq.\ (\ref{eq200}) will only take the two values
$f\rund{0}$ and  
$f\rund{1}$, so obviously $f\rund{0}\neq f\rund{1}$ is required,
because otherwise $r$ would become independent of the galaxy
distribution and useless for the correlation analysis. From these
statements it follows that we can find two constants $a$ and $b$ such
that, for $z_i=0$ or $1$,
\bdm
   \wt{f}\rund{z_i}:=a\cdot f\rund{z_i}+b=\frac{z_i}{N}\ \ ,
   \ \ \ i=1,\ldots,n\ ,
\edm
which in turn results in a linear transformation of the correlation
coefficient 
\bdm
   \wt{r}:=\sum_{i=1}^n g\rund{\r_i}\,\wt{f}\rund{z_i}
   =a\cdot r+b'\ \ ,
\edm
where
\bdm 
   b'=b\cdot\sum_{i=1}^n g\rund{\r_i}\ .
\edm

A substitution of $r$ with $\wt{r}$ in Eqs.\ (\ref{eq150})
and (\ref{eq160}) yields $\wt{Q}_{1,2}={\rm sign}\rund{a}\cdot
Q_{1,2}$ and $\vert\wt{Q}_{1,2}\vert=\abs{Q_{1,2}}$. This shows that
the linear transformation $f\rightarrow\wt{f}$ does not influence the
maximisation of $\abs{Q_1}$ or $\abs{Q_2}$, our means for optimizing
the correlation test. If $n$ is large enough (i.e. $z_i=0,1$) we can
therefore choose $f\rund{z_i}=z_i/N$ without loss of generality, so
we get
\bdm
   r:=\frac{1}{N}\sum_{i=1}^n z_i\,g\rund{\r_i}\ \ .
\edm
Furthermore, if the radial positions of the $N$ galaxies within
$\eck{\f_{\rm in},\f_{\rm out}}$ are denoted with $\f_j$,
$j=1,\ldots,N$, it is clear that with increasing $n$ the radii of
sub-rings $\eck{\r_i,\r_{i+1}}$ containing one of the galaxies
approach the values $\f_j$. As $z_i=0$ for all the other sub-rings, we 
can define
\beq\label{eq300}
   r_g\rund{\f_1,\ldots,\f_N} :=
   \lim_{n\to\infty}\frac{1}{N}\sum_{i=1}^n z_i\,g\rund{\r_i}=
   \frac{1}{N}\sum_{j=1}^N g\rund{\f_j}\ \ .
\eeq

Note that, by taking the limit $n\to\infty$, we finally got rid of the
arbitrary number $n$ of sub-rings within the galaxy field under
consideration. This is a first advantage of the `weighted average'
correlation test compared to Spearman's rank test, which was applied
in earlier investigations (e.g. BS) and required a binning of the
data.

In the following, we will assume the galaxies to be distributed
independently of one another. That means the galaxy distribution
within $\eck{\f_{\rm in},\f_{\rm out}}$ can be described by a
one-dimensional radial probability density $p\rund{\f}$, which
defines the probability $p\rund{\f}\de\f$ for a single
galaxy to be found at some position between two given angular radii
$\f$ and $\f+\de\f$. Because of galaxy-galaxy correlations our
assumption is a simplification and does not hold in general. In
Sect.\ \ref{qsoiras} we argue that it should nevertheless be applicable
in this analysis. 

One approach to statistically prove an association between quasars and
galaxies is to start with the `null-hypothesis' of no association and
then try to reject it with the help of the correlation coefficient 
(\ref{eq100}). Normally the null-hypothesis will be characterised by 
a radial probability density $p_{\rm o}\rund{\f}\sim \f$,
corresponding to a Poissonian galaxy distribution. However, if the
galaxy fields around the individual quasars that we merge to form the
total galaxy field do not all have the same angular radius or are
irregularly shaped, $p_{\rm o}\rund{\f}$ may look quite differently.
Therefore, we will write
\bdm
   p_{\rm o}\rund{\f} =: c_{\rm o}\cdot G\rund{\f}\ \ ,
\edm
using a geometrical factor $G\rund{\f}\ge 0$ and a normalisation
constant $c_{\rm o}>0$ to meet the condition
\bdm
   \int_{\f_{\rm in}}^{\f_{\rm out}}
   p_{\rm o}\rund{\f} \de\f=
   c_{\rm o}\cdot\int_{\f_{\rm in}}^{\f_{\rm out}}
   G\rund{\f} \de\f = 1\ \ .
\edm

The theory of gravitational lenses, together with some cosmological
model, can give quantitative predictions about the expected angular 
quasar-galaxy two-point correlation function $\xi_{\rm qg}\rund{\f}$
(cf. Bartelmann \cite{msb95}). For galaxy positions which are
independent of one another, this implies a one-dimensional
probability density $p_{\rm a}\rund{\f}$, which we will write in the 
form
\bdm
   p_{\rm a}\rund{\f} =: c_{\rm a}\rund{\f}\cdot G\rund{\f}\ \ ,
\edm
where $c_{\rm a}\rund{\f}$ is related to $\xi_{\rm qg}\rund{\f}$
via the expression
\bdm
   c_{\rm a}\rund{\f} = C\cdot\eck{1+\xi_{\rm qg}\rund{\f}}\ \ .
\edm
As before, normalisation is required, i.e.
\bdm
   \int_{\f_{\rm in}}^{\f_{\rm out}}
   p_{\rm a}\rund{\f} \de\f=
   C\cdot\int_{\f_{\rm in}}^{\f_{\rm out}} 
   \eck{1+\xi_{\rm qg}\rund{\f}} 
   G\rund{\f} \de\f = 1\ \ .
\edm

Given an estimate of $\xi_{\rm qg}\rund{\f}$ from theory we suspect
that the true distribution of galaxies might be described by $p_{\rm
a}\rund{\f}$, so we want the correlation coefficient (\ref{eq300}) to
be optimised for a distinction between $p_{\rm o}\rund{\f}$,
representing the null-hypothesis, and $p_{\rm a}\rund{\f}$. 
Applying definition (\ref{eq150}) this can be achieved by choosing
the weight function $g\rund{\f}$ such that the global maximum of
$\abs{Q_1}$ is reached. A variational calculation, shown in
Appendix~A.1, yields Eq.\ (\ref{eq40}),
\beq\label{eq400}
   g\rund{\f} = a\,\xi_{\rm qg}\rund{\f}+b\ \ ,
\eeq
with arbitrary ($a\ne 0$) constants $a$ and $b$. Similarly, one finds
\beq\label{eq500}
   g\rund{\f} = \frac{a'}{1+\xi_{\rm qg}\rund{\f}}+b'
\eeq
when maximising $\abs{Q_2}$. Such an optimisation of the correlation
test is most important if $p_{\rm o}\rund{\f}\approx p_{\rm
a}\rund{\f}$, i.e. $\xi_{\rm qg}\rund{\f}\ll 1$, because in this case
it will be most difficult to rule out one of the possibilities. In the
interesting regime of $\xi_{\rm qg}\ll 1$, however, the expansion
$\rund{1+\xi}^{-1}\approx 1-\xi$ implies Eqs.\ (\ref{eq400}) and
(\ref{eq500}) to be nearly equivalent. Moreover, Appendix~A.2
gives the proof that the weight function (\ref{eq400}) is also a
stationary point of the mean error level 
$\ave{P_{\rm o}\rund{r_{g}\ge R}}$, as long as 
$\xi_{\rm qg}\rund{\f}\ll 1$.   
\subsection{Numerically derived quasar-galaxy two-point correlation 
   function}
\label{numqgcorr}
\begin{figure}
   \centerline{\psfig{figure=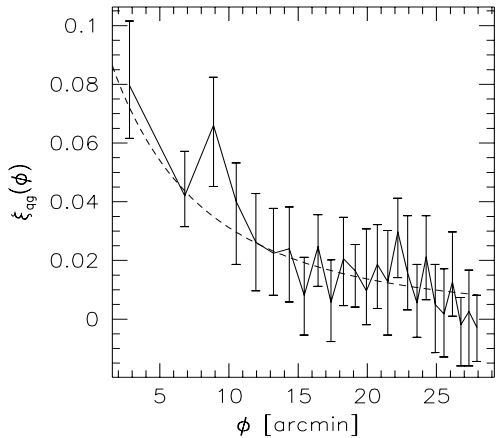,height=7.0cm,width=7.0cm}}
   \caption[]{Solid line: Quasar-galaxy two-point correlation function
      from a CDM Einstein-de~Sitter cosmological model with Hubble
      constant $H_{\rm o}=100\,{\rm km/(s\,Mpc)}$. Dashed line:
      Approximation according to (\protect\ref{eq600}).}
   \label{fig2}
\end{figure}

The theoretically expected angular two-point correlation function
between background quasars and foreground galaxies has been derived by
Bartelmann (\cite{msb95}) for cold dark matter (CDM) and hot dark
matter (HDM) Einstein-de~Sitter cosmological models with linearly
evolved perturbation spectra. For the following analyses we will pick
out the quasar-galaxy correlation function $\xi_{\rm qg}\rund{\f}$
which Bartelmann extracted from a numerical simulation\footnote{
   For details, see the original paper.
} of a CDM universe with $H_{\rm o}=100\,{\rm km/(s\,Mpc)}$ taking
into account galaxies up to $21^{\rm st}$ magnitude. Figure \ref{fig2}
shows plots of $\xi_{\rm qg}\rund{\f}$ from the numerical simulation
(solid line) and of the approximation 
\beq\label{eq600}
   \xi_{\rm qg}\rund{\f}\approx\xi'\rund{\f}:=
   \alpha\,\rund{\f_{\rm o}+\f/{\rm deg}}^{-2.4}\ \ ,
\eeq
with $\alpha=0.0036$ and $\f_{\rm o}=0.24$ (dashed line). Whereas
Bartelmann's correlation functions are all normalised such that
\bdm
   \int_0^{\infty} \xi_{\rm qg}\rund{\f} \f\,\de\f = 0\ \ ,
\edm
this is obviously not true for $\xi'\rund{\f}$, because
$\xi'\rund{\f}>0$ on $\f \in \eck{0,\infty}$. In this sense
$\xi'\rund{\f}$ is not a valid fit to $\xi_{\rm qg}\rund{\f}$.
Nonetheless the approximation does resemble three important features
of the correlation function: $\xi'\rund{\f}$ is steep for small $\f$,
flattens at $\f\approx\f_{\rm o}$ and is very shallow for large values
of $\f$. We will therefore use $\xi'\rund{\f}$ to derive an
approximation to the optimum weight function. To allow for different
values $H_{\rm o}=h\cdot 100\,{\rm km/(s\,Mpc)}$ of the Hubble
constant we have to multiply\footnote{
   This can be concluded from Eqs. (2.12), (2.32) and (2.33) of
   Bartelmann (\cite{msb95}): Observing that $k\sim h$ and 
   $k_0\sim h$ (see e.g. Padmanabhan, \cite{pad}) one finds
   $\xi_{\rm qg}\rund{\f;h=h'}= 
   C\rund{h'}\,\xi_{\rm qg}\rund{h'\,\f;h=1}$. 
} $\f$ by the dimensionless parameter $h$. Setting $a=1/\alpha$ and
$b=0$ we thus obtain from Eq.\ (\ref{eq40})
\beq\label{eq650}
   g\rund{\f}=\rund{0.24+h\,\f/{\rm deg}}^{-2.4}\ \ .
\eeq

%
%
%
%
%
%
%
\section{Numerical simulations}
\label{numsim}
Before applying the newly introduced weighted-average correlation test
to observational data we want to show the results of some numerical
simulations. They have been performed in order to check whether our
method is more sensitive than the Spearman's rank-order test, 
and to investigate the influence of variations of the
weight function.
\subsection{Overview}
In principle, we proceed as follows: First, a large number (we use
$10^7$) of circular galaxy fields is generated, each of them
consisting of $N$ independently and randomly chosen galaxy 
positions. We interpret them as a statistical sample of total
galaxy fields in the case of the null-hypothesis of no quasar-galaxy
correlation and use them to derive the distribution $P_{\rm
o}\rund{r_g\ge R}$. A second set (size: $10^6$) of random
fields is synthesized to model observed galaxy fields with an assumed
association of quasars and galaxies. Again, the coordinates are drawn
independently, but this time according to some radial probability
density profile $p_{\rm a}\rund{\f}$, which is meant to describe the
correlation. For a given weight function $g\rund{\f}$ we calculate
$\ave{r}_{\rm a}$ as well as $\ave{P_{\rm o}\rund{r_g\ge r}}_{\rm a}$
from the individual values $r$ of the correlation coefficient $r_g$.
We use these quantities to analyse the sensitivity of our
weighted-average correlation test as described below.

Both the error level $P_{\rm o}\rund{r_g\ge\ave{r}}$ of the
mean correlation coefficient and the mean error level $\ave{P_{\rm
o}\rund{r_g\ge r}}$ quantify the distinction between our two
simulated samples: very low values (or values near unity) signify a
good distinction, values around $0.5$ a bad distinction. 
The effect of deviations of the weight function from its optimum
$g\rund{\f}$ can be examined by defining a parameterized weight
function $g\rund{\f,\e}$ such that $g\rund{\f,\e=\e_{\rm o}}\equiv
g\rund{\f}$ for some $\e=\e_{\rm o}$ and repeating the simulation for
a whole set of values for $\e$.    

Additionally we subject each of the galaxy fields of the second
sample to Spearman's rank-order test to obtain the corresponding error
level of the mean correlation coefficient and the mean error
level. That allows for a direct comparison between the
weighted-average method and Spearman's rank-order test, a short
description of which is given below:  
\begin{itemize}
   \item The circular galaxy field is divided into $n=25$ rings of
   equal area.
   The number $25$ has been adopted from the Bartelmann \& Schneider
   (\cite{msb94}) analysis of quasar-galaxy associations.
   \item The number of galaxies within each ring is determined and
   ranked, yielding a scheme of ranks $\rund{{\cal R}_1^{\rm a},
   \ldots, {\cal R}_{25}^{\rm a}}$ with ${\cal R}_i^{\rm a} \in 
   \wave{1,\ldots,25}$.
   \item The distance of the rings from the center is ranked in
   descending order, i.e. rings closer to the center are ranked
   higher. The result is a second rank scheme 
   $\rund{{\cal R}_1^{\rm b},\ldots,{\cal R}_{25}^{\rm b}}= 
   \rund{25,24,\ldots,1}$.
   \item Spearman's rank-order correlation coefficient $r_{\rm s}$ is
   calculated from its definition as the linear correlation
   coefficient of the rank schemes,
   \bdm
      r_{\rm s}:=\frac{\sum_{i=1}^n 
      \rund{{\cal R}_i^{\rm a}-\ol{\cal R}^{\rm a}}
      \rund{{\cal R}_i^{\rm b}-\ol{\cal R}^{\rm b}}}
      {\sqrt{\sum_{i=1}^n
      \rund{{\cal R}_i^{\rm a}-\ol{\cal R}^{\rm a}}^2}
      \sqrt{\sum_{i=1}^n
      \rund{{\cal R}_i^{\rm b}-\ol{\cal R}^{\rm b}}^2}}\ \ ,
   \edm
   where 
   \bdm
      \ol{\cal R}^{\rm a,b}:=
      \rund{1/n}\sum_{i=1}^n {\cal R}_i^{\rm a,b}\ \ .
   \edm
   Taking into account that the rank schemes are always permutations
   of $\wave{1,\ldots,n}$ this expression can be simplified to
   \bdm
      r_{\rm s}=1-\frac{6\cdot\sum_{i=1}^n 
      \rund{{\cal R}_i^{\rm a}-{\cal R}^{\rm b}}^2}
      {n\rund{n^2-1}}\ \ .
   \edm 
   \item If the number of galaxies within each ring is independent of
   the distance of the ring from the center and independent of the
   numbers of galaxies within the other rings, then the rank schemes
   $\wave{{\cal R}_i^{\rm a}}$ and $\wave{{\cal R}_i^{\rm b}}$ are
   random permutations of each other. In this case the distribution of
   \bdm
      t:=r_{\rm s}\eck{\rund{n-2}/\rund{1-r_{\rm s}^2}}^{1/2}
   \edm
   for $n>10$ is excellently approximated by a Student-t-distribution.
   Therefore, the error level of Spearman's rank-order test is given
   by 
   \bdm
      P\rund{r_{\rm s}\ge R}=\left\{\ 
      \begin{array}{ll}
         \frac{1}{2}I_{1-R^2}\rund{\frac{n-2}{2},\frac{1}{2}} &
         \mbox{for $R\ge 0$}\ \ ,\\
         1-\frac{1}{2}I_{1-R^2}\rund{\frac{n-2}{2},\frac{1}{2}} &
         \mbox{for $R< 0$}\ \ ,
      \end{array}
      \right.
   \edm
   with $I_z\rund{a,b}$ denoting the incomplete beta function
   \bdm
      I_z\rund{a,b}:=\frac{\int_0^z s^{a-1}\rund{1-s}^{b-1}\,\de s}
      {\int_0^1 s^{a-1}\rund{1-s}^{b-1}\,\de s}\ \ .
   \edm
\end{itemize}
\subsection{Results}
%
%
%
\begin{figure*}
   \centerline{\psfig{figure=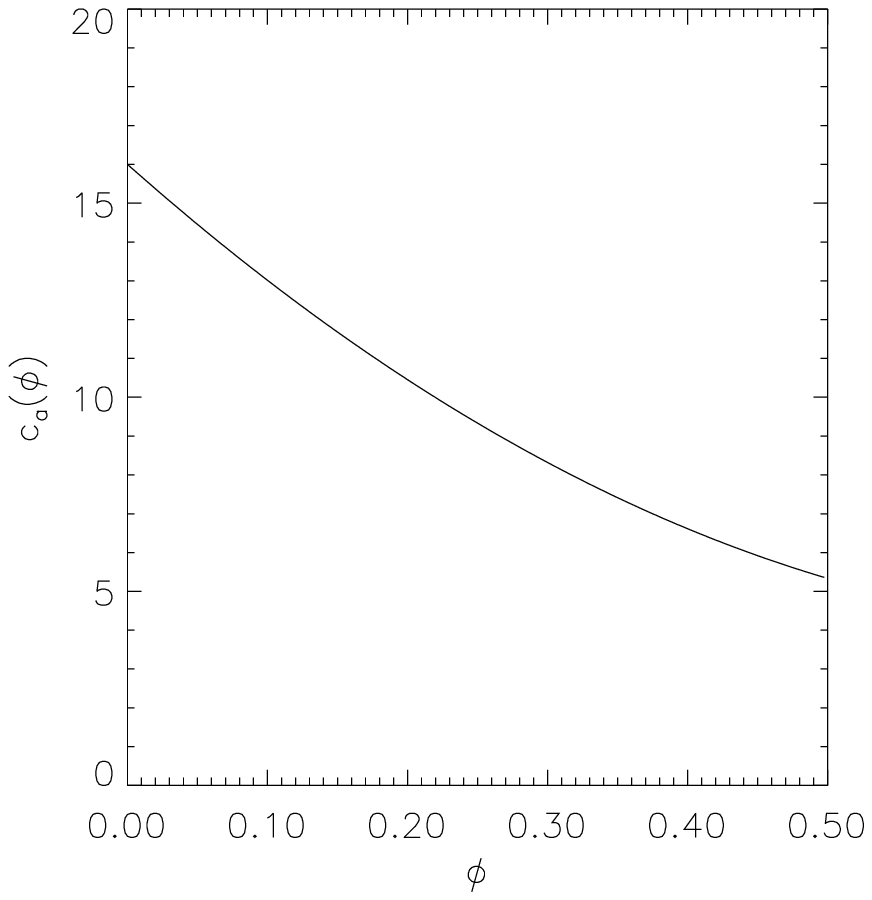,height=7.0cm,width=7.0cm}
   \hfill\psfig{figure=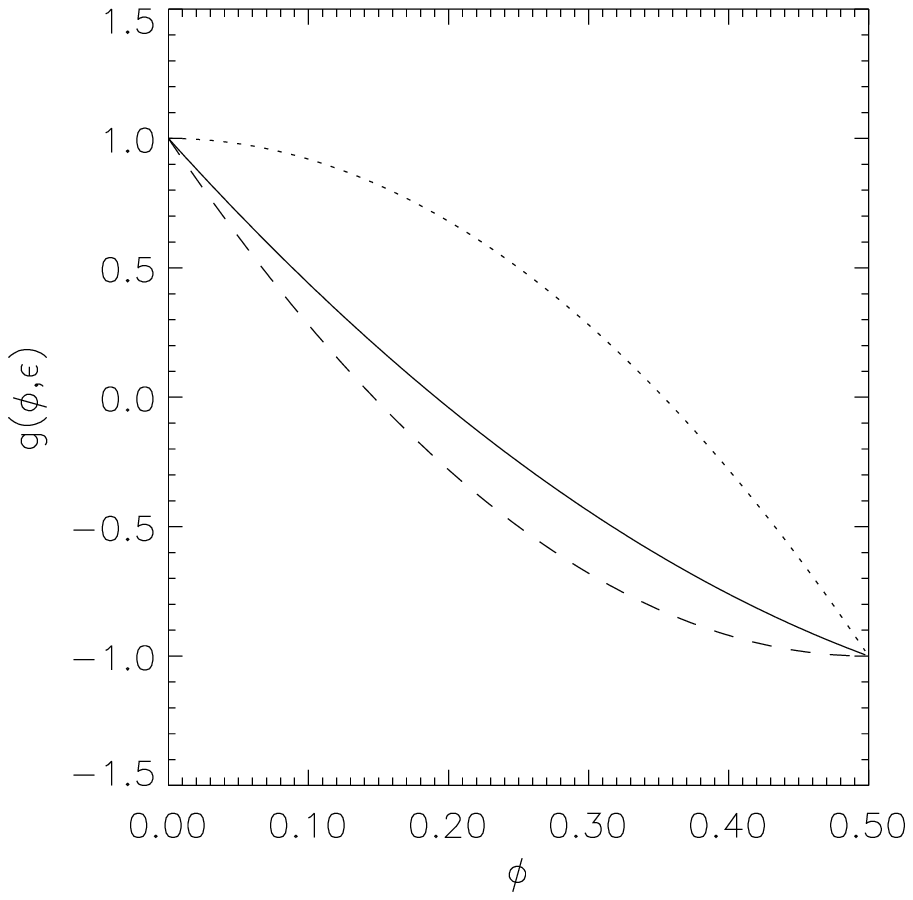,height=7.0cm,width=7.0cm}\hfill}
   \caption[]{Left panel: ``Correlation profile'' $c_{\rm a}\rund{\f}$ 
      according to expression~(\protect\ref{eq700}). Right panel:
      Parameterized weight function $g\rund{\f,\e}$ as defined
      in (\protect\ref{eq800}) for $\e=0.75$ (solid line), $\e=0$ 
      (dotted line) and $\e=1$ (dashed line).}
   \label{fig3}
\end{figure*}
%
%
In the following, all angles will be measured in degrees and treated
as dimensionless quantities.
On circular galaxy fields the null-hypothesis corresponds to a 
probability density of galaxy positions
\bdm 
   p_{\rm o}\rund{\f}=c_{\rm o}\,G\rund{\f}\sim\f\ \ . 
\edm
Let us fix the radius of the simulated galaxy fields to $0.5$
(degrees) and set
\bdm
   G\rund{\f}:=\f\ \ ,
\edm
which requires $c_{\rm o}=8$ in order to assure the normalisation of 
$p_{\rm o}\rund{\f}$. For our first series of simulations we
arbitrarily choose
\beq\label{eq700}
   p_{\rm a}\rund{\f}=c_{\rm a}\rund{\f}\,G\rund{\f}:=
   \rund{\frac{64}{3}\f^2-32\f+16}\cdot G\rund{\f}
\eeq
to describe a hypothetical quasar-galaxy correlation on the galaxy
fields (Fig.\ \ref{fig3}, left panel). Moreover we define a
parameterized weight function (Fig.\ \ref{fig3}, right panel)
\beq\label{eq800}
   g\rund{\f,\e}:=\rund{16\e-8}\f^2-8\e\f+1\ \ ,
\eeq
that satisfies the condition (\ref{eq40}) for an optimum weight
function if $\e=0.75$. From these expressions we can derive the mean
correlation coefficient
\beq\label{eq900}
   \ave{r} := \int\!\!\cdots\!\!\int\eck{\frac{1}{N}\sum_{i=1}^N
   g\rund{\f_i,\e}}\prod_{j=1}^N \eck{p_{\rm a}\rund{\f_j}\,
   \de\f_j} = \int_0^{0.5} g\rund{\f,\e}\, 
   p_{\rm a}\rund{\f}\,\de\f = \frac{7-32\e}{45}\ \ .
\eeq

\begin{figure*}
   \centerline{
   \hfill\psfig{figure=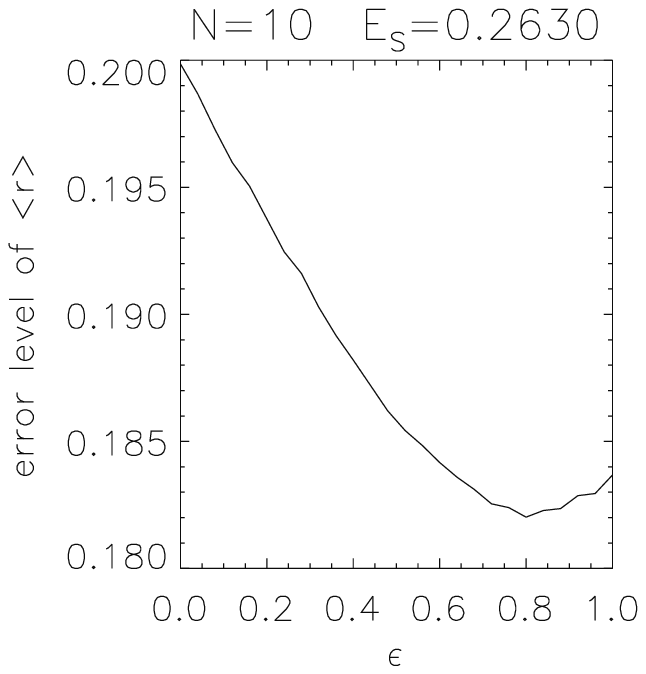,height=7.0cm,width=7.0cm}
   \hfill\psfig{figure=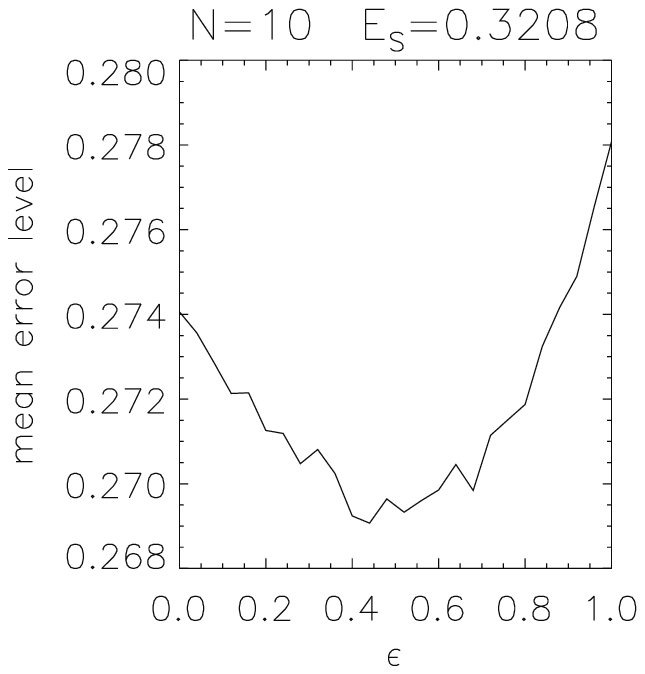,height=7.0cm,width=7.0cm}\hfill}
   \vspace{-0.2cm}
   \centerline{
   \hfill\psfig{figure=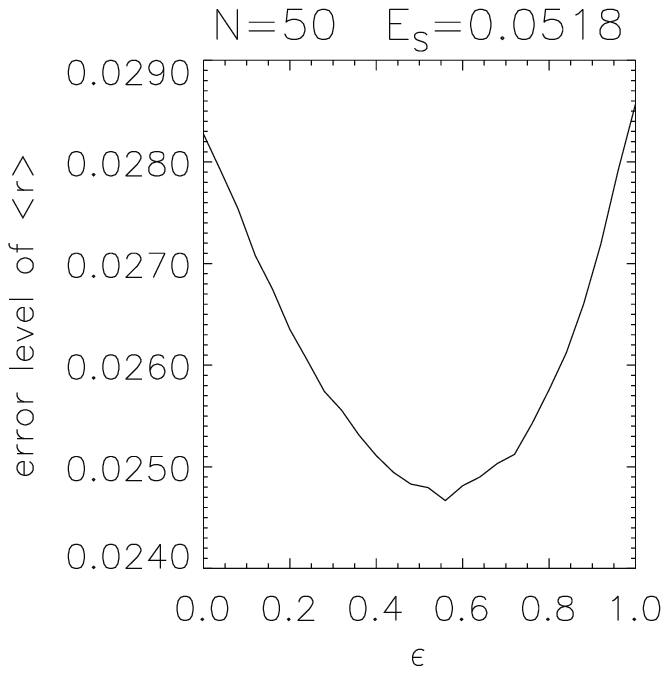,height=7.0cm,width=7.0cm}
   \hfill\psfig{figure=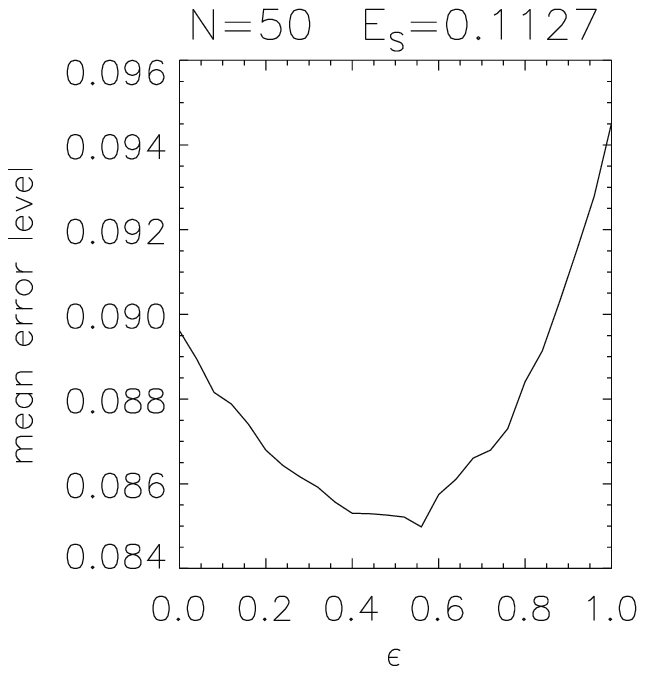,height=7.0cm,width=7.0cm}\hfill}
   \vspace{-0.2cm}
   \centerline{
   \hfill\psfig{figure=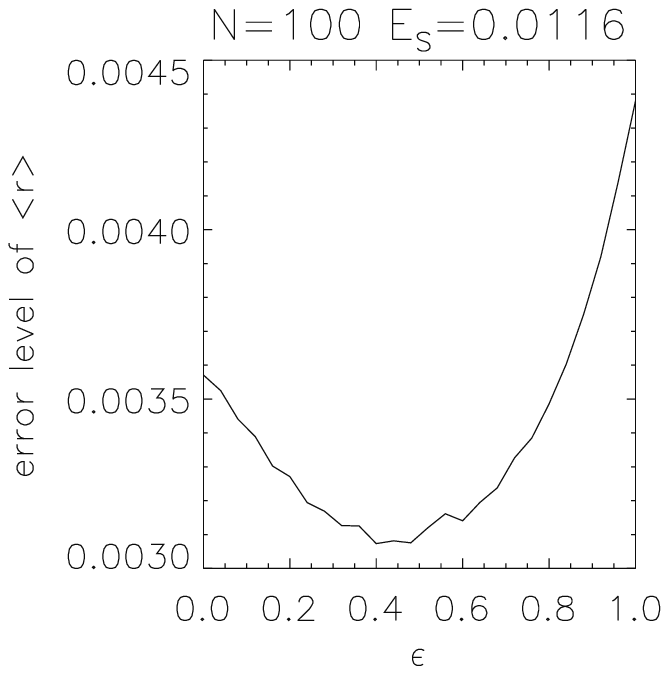,height=7.0cm,width=7.0cm}
   \hfill\psfig{figure=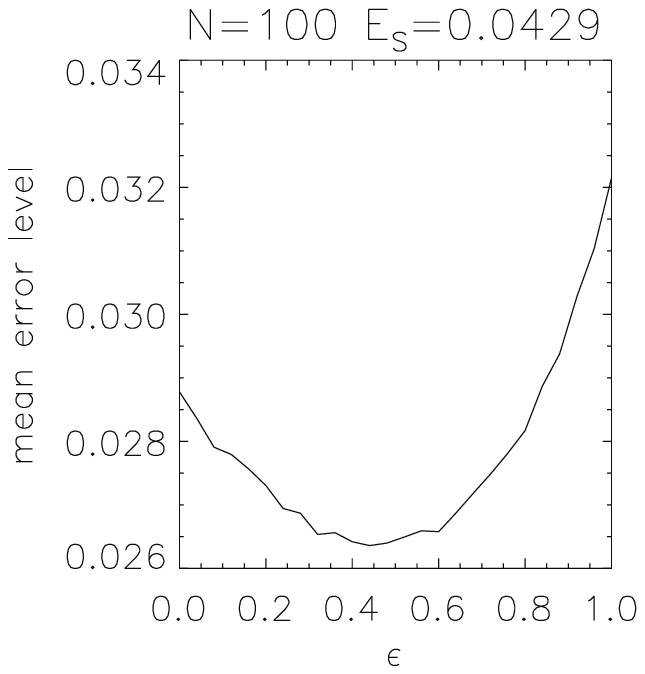,height=7.0cm,width=7.0cm}\hfill}
   \vspace{-0.2cm}
   \caption[]{Error level of the mean correlation coefficient (left
   column) and mean error level (right column) for the
   weighted-average 
   correlation test as a function of the parameter $\e$ of the
   weight function (\protect\ref{eq800}). The heading of each panel
   displays the corresponding results from Spearman's rank-order test
   ($E_{\rm S}$) and the number of galaxies $N$.}
   \label{fig4}
\end{figure*}

The outcome of simulations for different values of $N$ carried out as
explained in the preceding section is summarized in
Fig.~\ref{fig4}. The individual plots visualize the error level of the
mean correlation coefficient or the mean error level, respectively, as
a function of the weight-function parameter $\e$. In the heading of
each panel one can find, apart from the number of galaxies $N$, the
results obtained from Spearman's rank-order test (which are, of
course, independent of $\e$), denoted by $E_{\rm S}$.
Two important facts can be seen in Fig.\ \ref{fig4}:
\begin{itemize}
   \item Both the error level of the mean correlation coefficient and
   the mean error level do indeed attain a minimum for a weight function
   near the expected optimum $g\rund{\f,\e=0.75}$. However, the exact
   position of the minimum is slightly displaced with respect to the
   predicted one.    
   \item Regarding the error level of the mean correlation coefficient
   as well as regarding the mean correlation coefficient, the
   weighted-average correlation test yields a more significant
   distinction between the two samples of galaxy fields than
   Spearman's rank-order 
   test. This remains true even if the weight function deviates 
   significantly from its optimum.     
\end{itemize} 

%
%
%
%

\begin{figure*}
   \centerline{\psfig{figure=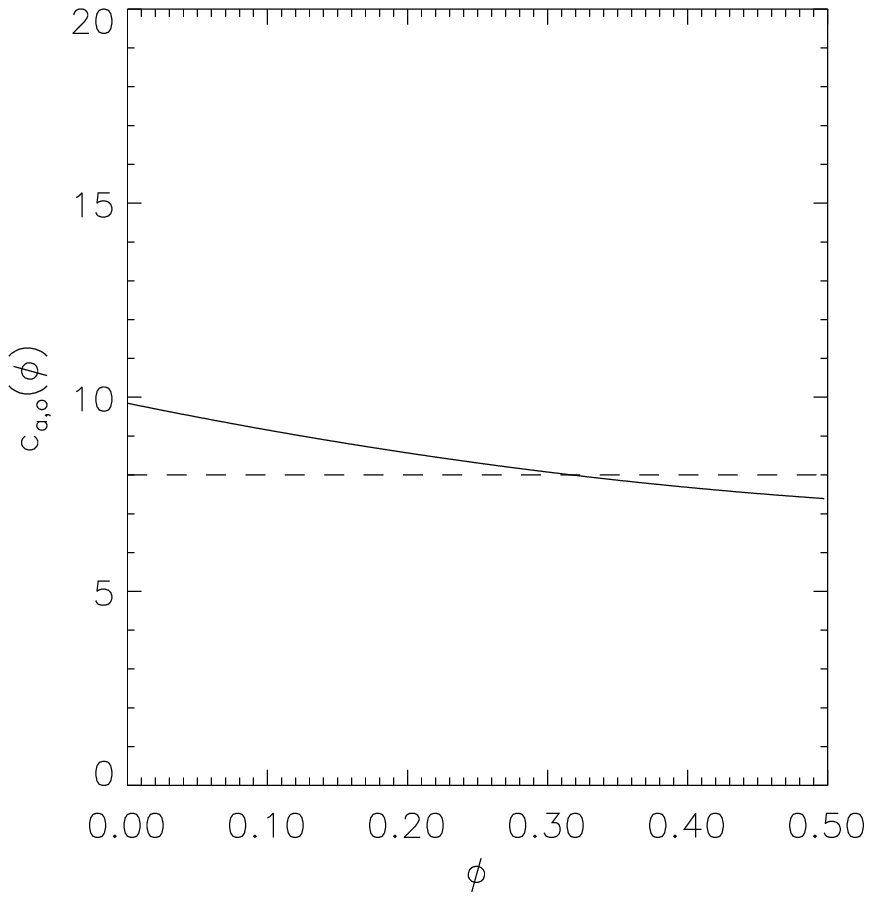,height=7.0cm,width=7.0cm}
   \hfill\psfig{figure=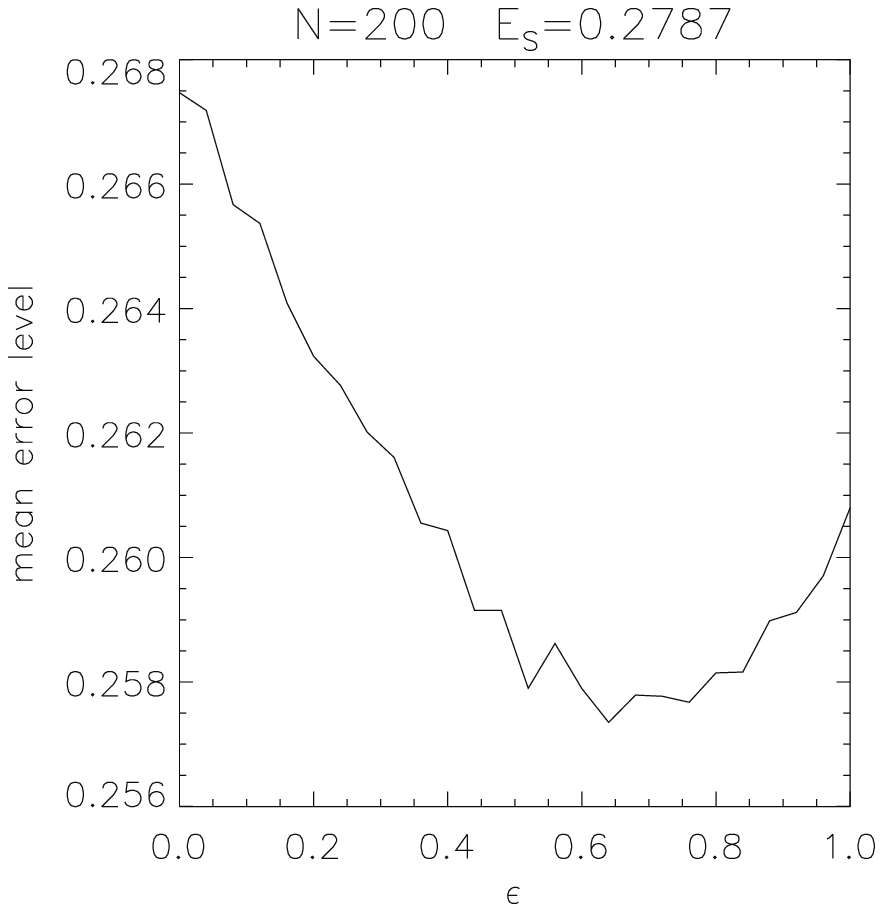,height=7.0cm,width=7.0cm}\hfill}
   \caption[]{Left panel: ``Correlation profile'' $c_{\rm a}\rund{\f}$ 
      according to Eq.\ (\protect\ref{eq1000}) (solid line) and 
      constant value $c_{\rm o}=8$ (dashed line). Right panel: Mean
      error level for the weighted-average correlation test as a
      function of the parameter $\e$ of the weight function
      (\protect\ref{eq800}). The number of galaxies is $N=200$. As in
      Fig.\ \protect\ref{fig4} $E_{\rm S}$ denotes the corresponding
      result obtained from Spearman's rank-order test.}
   \label{fig5}
\end{figure*}

As to the stated displacement of the minimum of the mean error level we
refer to Appendix~A.2 where the optimum weight function
(\ref{eq40}) is shown to approach a stationary point of the
mean error level if $p_{\rm a}\rund{\f}$ comes close enough to $p_{\rm
o}\rund{\f}$. This
suggests that the real position of the minimum should be closer to the
predicted one if the difference between $p_{\rm a}\rund{\f}$ and
$p_{\rm o}\rund{\f}$ is smaller. In order to verify this, we performed
a second series of simulations (in this case the null-hypothesis was
modelled using only $5\cdot 10^6$ random fields), now setting (cf.\
Fig.\ \ref{fig5}, left panel)
\beq\label{eq1000}
   p_{\rm a}\rund{\f}=c_{\rm a}\rund{\f}\,G\rund{\f}:=
   \frac{32}{13}\rund{2\f^2-3\f+4}\cdot G\rund{\f}\ \ .
\eeq
The right panel of Fig.\ \ref{fig5} displays the mean error level for
$N=200$ galaxies in the total galaxy field. Clearly, the minimum is
located closer to the expected position than before. We can also see
from Fig.\ \ref{fig5} that Spearman's rank-order test is actually not
too bad compared to our new test. It should also be noted that the
gain in sensitivity of our new test is obtained by providing more a
priori information to the test, namely the shape of the expected
two-point correlation function. On the other hand, the correlation
function chosen here is particularly favourable for Spearman's rank-
order test, since it can be approximated nearly as a linear function
of the angular radius on the scales considered.

%
%
%
%
%
%
%
\section{Data}
\label{qsoiras}
Now that the weighted average as a method to detect a possible
quasar-galaxy associations has been tested numerically, we are going
to reanalyze the correlations between 1Jy quasars and IRAS galaxies
reported in BS. Because of the
high significance of their results it seems promising to try to find
additional information such as a correlation scale or amplitude.
\subsection{Sample selection}
Our investigation is based on exactly the same data set as was used in
BS; the reader is referred to this paper for details.

The positions and photometric data of the galaxies are taken from the
Infrared Astronomical Satellite (IRAS) Faint Source Catalog, applying 
the criterion
\bdm
   S_{60}^2 \ge S_{12}\,S_{25}
\edm
to identify a source as a likely galaxy, where $S_n$ denotes the flux
at $n$ micron. To exclude very faint as well as very strong nearby
sources the sample is then restricted to objects within the range
\bdm
   0.3\,\mbox{Jy} \le S_{60} \le 1\,\mbox{Jy}\ \ .
\edm

A sample of quasars is provided by the optically identified fraction
of the 1Jy catalog. This catalog contains bright extragalactic radio
sources with a 5-GHz flux of above 1 Jy (K\"uhr et al.\
\cite{kuehr}, Stickel et al.\ \cite{sti93}, Stickel \& K\"uhr
\cite{sti93a}, \cite{sti93b}). From the 426 radio sources with known
redshift we select different subsamples, each of them characterized by
a lower and an upper limit $z_{\rm min}$ and $z_{\rm max}$ of the
redshift as well as an upper limit $m_{\rm max}$ of the apparent
magnitude. 

To perform the statistical analysis we extract from the
IRAS catalog a ring-like galaxy field of radii $\f_{\rm in}$ and
$\f_{\rm out}$ around the position of each of the selected quasars.
Merging them as explained in Sec.\ \ref{qgcorr} we end up with a total
ring-like galaxy field of $N$ galaxies, which can be subjected to the 
weighted-average correlation test. It has already been pointed out
that we assume the galaxies to be distributed independently of one
another. In particular this assumption enters the derivation of the
optimum weight function and is fundamental for the calculation of any
error level given in this paper, because our null-hypothesis is a
Poissonian distribution of galaxies. However, the number density of
IRAS galaxies is of the order of 1 per square-degree, and the
individual galaxy fields are taken from very different positions on
the sky. As a consequence, galaxy-galaxy correlations on the merged
total field should be negligible. 
\subsection{Correlation analysis}
In order to allow for a direct comparison of our results with those
obtained in BS using Spearman's rank-order test,
we first adopt the value of $\f_{\rm out}=28.21$ arcmin. This radius
had been chosen such that the area $\pi\f_{\rm out}^2$ around each of
the quasars was $2500$ arcmin$^2$, in agreement with an earlier
investigation of 1Jy-sources and Lick galaxies by Fugmann
(\cite{fug90}).
 
As the IRAS catalog contains a few sources which are located within a
few arcseconds from the position of one of the 1Jy-quasars (cf.\
BS) and can therefore possibly be
identified with the corresponding quasars, we remove a small circle
from the center of our galaxy fields by setting $\f_{\rm in}=10$
arcsec.

Tables \ref{ta1} and \ref{ta2} summarize the results of the
correlation tests. The different columns present the following
information: $z_{\rm min}$ denotes the minimum redshift, $z_{\rm max}$
the maximum redshift, $m_{\rm max}$ the maximum apparent magnitude of
the selected quasar subsample and $N_{\rm Q}$ the number of quasars
within this subsample. The total number of galaxies is $N$ and
the two remaining columns display the error levels $\e_{\rm wa}$ and
$\e_{\rm sp}$ in percent as obtained from the weighted-average and
Spearman's rank-order tests, respectively. The former is derived from
a numerical simulation of $10^6$ random (Poissonian) galaxy fields,
the latter is taken from BS.
In Table \ref{ta1}, where no value of $z_{\rm max}$ is given, the
subsamples do not have an upper limit of the redshift, i.e. formally
it is $z_{\rm max}=\infty$.

%
%
\begin{table}
   \caption[]
      {Results of the correlation tests between 1Jy quasars and
      IRAS galaxies. The error level (in \%) as obtained from the
      weighted average is denoted by $\e_{\rm wa}$, whereas
      $\e_{\rm sp}$ is the error level from the Spearman's rank test
      as reported in BS. The remaining symbols are $z_{\rm min}$ and 
      $m_{\rm max}$ for the minimum redshift and maximum apparent 
      magnitude of the quasar subsample, $N_{\rm Q}$ for the number of
      quasars, $N$ for the number of galaxies and $\f_{\rm out}$
      denoting the outer radius of the galaxy fields.}
   \label{ta1}
   \begin{center}
      \begin{tabular}{|rr|r|r|r||r|}
         \hline
          & & & \multicolumn{3}{c|}{$\f_{\rm out}=28.21'$}\\
         \cline{4-6}
         $z_{\rm min}$ & $m_{\rm max}$ & $N_{\rm Q}$ & $N$ & 
         $\e_{\rm wa}$ & $\e_{\rm sp}$\\
         \hline\hline
         0.50 & 21.00 & 238 & 159 &   .7 & 23.5 \\
         0.50 & 20.00 & 218 & 152 &   .4 & 14.1 \\
         0.50 & 19.00 & 192 & 141 &  1.3 & 20.5 \\
         0.50 & 18.75 & 169 & 127 &  2.3 & 30.5 \\
         0.50 & 18.50 & 161 & 114 &  2.8 & 49.3 \\
         0.50 & 18.25 & 123 &  89 & 19.5 & 30.9 \\
         0.50 & 18.00 & 114 &  82 & 34.5 & 46.2 \\
         \hline			          
         0.75 & 21.00 & 179 & 119 &  2.7 & 15.9 \\
         0.75 & 20.00 & 166 & 113 &  1.4 & 16.0 \\
         0.75 & 19.00 & 144 & 105 &  3.7 & 16.2 \\
         0.75 & 18.75 & 126 &  93 &  8.4 & 24.4 \\
         0.75 & 18.50 & 120 &  84 &  9.4 & 29.0 \\
         0.75 & 18.25 &  87 &  64 & 42.1 & 37.7 \\
         0.75 & 18.00 &  80 &  58 & 67.4 & 38.2 \\
         \hline			         
         1.00 & 21.00 & 130 &  90 &  4.8 & 32.4 \\
         1.00 & 20.00 & 123 &  86 &  3.1 & 22.5 \\
         1.00 & 19.00 & 107 &  79 &  8.6 & 29.3 \\
         1.00 & 18.75 &  94 &  70 &  5.5 & 17.8 \\
         1.00 & 18.50 &  88 &  61 &  6.0 & 24.4 \\
         1.00 & 18.25 &  60 &  47 & 52.5 & 53.8 \\
         1.00 & 18.00 &  54 &  43 & 74.0 & 73.2 \\
         \hline			         
         1.25 & 21.00 &  97 &  65 &  1.0 &  2.8 \\
         1.25 & 20.00 &  93 &  63 &   .9 &  4.2 \\
         1.25 & 19.00 &  80 &  56 &  3.1 &  4.4 \\
         1.25 & 18.75 &  68 &  47 &  1.4 &  4.7 \\
         1.25 & 18.50 &  64 &  46 &  1.1 &  6.2 \\
         1.25 & 18.25 &  42 &  33 & 26.6 & 14.2 \\
         1.25 & 18.00 &  37 &  29 & 48.9 & 38.2 \\
         \hline			         
         1.50 & 21.00 &  59 &  33 &  9.5 &  0.9 \\
         1.50 & 20.00 &  56 &  33 &  9.5 &  0.2 \\
         1.50 & 19.00 &  46 &  30 &  5.6 &  0.7 \\
         1.50 & 18.75 &  36 &  23 &  3.8 &  0.4 \\
         1.50 & 18.50 &  34 &  22 &  2.9 &  0.3 \\
         1.50 & 18.25 &  20 &  14 &  7.9 &  0.2 \\
         1.50 & 18.00 &  18 &  14 &  7.9 &  1.2 \\
         \hline
      \end{tabular}
   \end{center}
\end{table} 
%
%

%
%
\begin{table}
   \caption[]
      {As table~\protect\ref{ta1} but for non-overlapping 
       redshift intervals of the quasar subsamples.} 
   \label{ta2}  
   \begin{center}
      \begin{tabular}{|rrr|r|r|r|}
         \hline
          & & & & \multicolumn{2}{c|}{$\f_{\rm out}=28.21'$}\\
         \cline{5-6}
         $z_{\rm min}$ & $z_{\rm max}$ & $m_{\rm max}$ & 
         $N_{\rm Q}$ & $N$ & $\e^{\rm wa}$ \\
         \hline\hline
         0.50 & 0.75 & 21.00 & 61 & 40 &  5.3 \\
         0.50 & 0.75 & 20.00 & 53 & 39 &  4.4 \\
         0.50 & 0.75 & 19.00 & 49 & 36 &  7.5 \\
         0.50 & 0.75 & 18.75 & 44 & 34 &  5.0 \\
         0.50 & 0.75 & 18.50 & 42 & 30 &  5.7 \\
         0.50 & 0.75 & 18.25 & 37 & 25 &  9.6 \\
         0.50 & 0.75 & 18.00 & 35 & 24 &  8.7 \\
         \hline 	                       
         0.75 & 1.00 & 21.00 & 49 & 29 & 14.1 \\
         0.75 & 1.00 & 20.00 & 43 & 27 & 10.6 \\
         0.75 & 1.00 & 19.00 & 37 & 26 &  9.9 \\
         0.75 & 1.00 & 18.75 & 32 & 23 & 49.2 \\
         0.75 & 1.00 & 18.50 & 32 & 23 & 49.2 \\
         0.75 & 1.00 & 18.25 & 27 & 17 & 28.7 \\
         0.75 & 1.00 & 18.00 & 26 & 15 & 38.9 \\
         \hline 	                        
         1.00 & 1.25 & 21.00 & 34 & 25 & 77.7 \\
         1.00 & 1.25 & 20.00 & 31 & 23 & 67.6 \\
         1.00 & 1.25 & 19.00 & 28 & 23 & 67.7 \\
         1.00 & 1.25 & 18.75 & 27 & 23 & 67.7 \\
         1.00 & 1.25 & 18.50 & 25 & 15 & 88.7 \\
         1.00 & 1.25 & 18.25 & 18 & 14 & 88.0 \\
         1.00 & 1.25 & 18.00 & 17 & 14 & 88.0 \\
         \hline 	                        
         1.25 & 1.50 & 21.00 & 38 & 32 &  2.2 \\
         1.25 & 1.50 & 20.00 & 37 & 30 &  1.6 \\
         1.25 & 1.50 & 19.00 & 34 & 26 & 13.3 \\
         1.25 & 1.50 & 18.75 & 32 & 24 &  8.1 \\
         1.25 & 1.50 & 18.50 & 30 & 24 &  8.1 \\
         1.25 & 1.50 & 18.25 & 22 & 19 & 67.7 \\
         1.25 & 1.50 & 18.00 & 19 & 15 & 96.3 \\
         \hline 	                        
         1.50 &$\infty$& 21.00 & 59 & 33 &  9.5 \\
         1.50 &$\infty$& 20.00 & 56 & 33 &  9.5 \\
         1.50 &$\infty$& 19.00 & 46 & 30 &  5.6 \\
         1.50 &$\infty$& 18.75 & 36 & 23 &  3.8 \\
         1.50 &$\infty$& 18.50 & 34 & 22 &  2.9 \\
         1.50 &$\infty$& 18.25 & 20 & 14 &  7.9 \\
         1.50 &$\infty$& 18.00 & 18 & 14 &  7.9 \\
         \hline
      \end{tabular}
   \end{center}
\end{table} 
%
%

A comparison of $\e_{\rm aw}$ and $\e_{\rm sp}$ reveals similar
properties of the error levels of the weighted average and Spearman's
rank test concerning their dependence on $z_{\rm min}$ and $m_{\rm
max}$. However, for subsamples with $z_{\rm min}<1.5$ the weighted
average very often yields strikingly lower values of the error level
(and thus statistically more significant results) than Spearman's rank
test. This effect becomes even more important if taking into account
that the $\e_{\rm sp}$ have been derived without removing the
mentioned IRAS counterparts of the 1Jy quasars from the galaxy
fields. Such counterparts can be found in the subsamples with 
$z_{\rm min}<1$, so the reported values of $\e_{\rm sp}$ for them are
probably too low.

As shown in BS, the IRAS catalog
probably extends to galaxy redshifts up to 1. As a consequence there
might be a spatial association between IRAS galaxies and 1Jy quasars
with a redshift below $z=1$ which, of course, could show up in our
correlation tests. Therefore, the quasar subsamples with $z_{\rm
min}\ge 1$ are the most interesting ones in the context of
gravitational lensing. Table \ref{ta2} gives a better insight into the
dependence of the correlations upon the quasar redshift, because the 
listed results are based on non-overlapping redshift bins. 
Surprisingly, we find an anticorrelation for $1.0 \le z < 1.25$, but
at a significance below $10\%$.

\begin{figure}
   \centerline{\psfig{figure=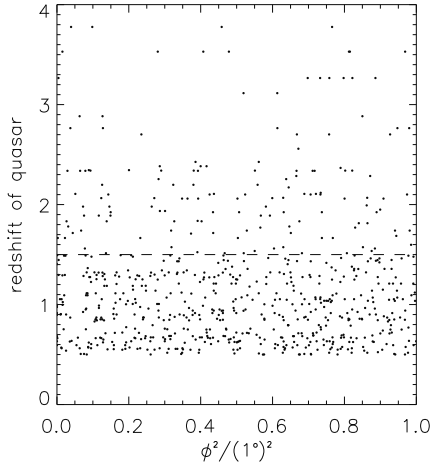,height=7.0cm,width=7.0cm}}
   \caption[]{Distribution of IRAS galaxies around 1Jy quasars of
   redshift $z\ge 0.5$ and apparent visual magnitude $m\le 21$.}
   \label{fig5.5}
\end{figure}

Now we want to investigate the question of how the results are
influenced by the choice of the outer radius $\f_{\rm out}$ of the
galaxy fields. As Spearman's rank test is sensitive to the overall
gradient of the galaxy number density over the field, $\f_{\rm out}$
must be adapted to the angular scale of the expected correlations. But
then a low error level can equally be induced by a galaxy overdensity
near the center or an underdensity in the outer regions of the field,
relative to the mean number density on large scales. Figure
\ref{fig5.5} shows the distribution of galaxies around 1Jy
quasars on fields of radius $1$ deg. Each dot corresponds to one
galaxy: The position along the horizontal axis indicates its distance
to the quasar, whereas on the vertical axis the quasar redshift can be
read off. The quadratic scaling of the horizontal axis assures that a
constant galaxy number density on the galaxy fields transforms to a 
constant number density of dots in the plot. Considering the
galaxy distribution around quasars of redshift $z>1.5$, i.e. the dots
above the dashed line, one finds that there seems to be a ``hole'' in
the range $0.15<\f^2/\rund{1 {\rm deg}}^2<0.25$ which is the outer
region of galaxy fields with $\f_{\rm out}\approx 30$ arcmin. This
deficit of galaxies could possibly yield a major contribution to the
low error levels recorded in Tables \ref{ta1} and \ref{ta2} for 
$z_{\rm min}=1.5$. To remove that contribution one would like to
increase $\f_{\rm out}$, thereby gaining information about the large
scale mean galaxy number density. On the other
hand this will decrease the overall density gradient once 
$\f_{\rm out}$ becomes larger than the correlation length scale and
therefore the significance of the result from Spearman's rank test
will decrease.

%
%
%
%

This problem can be overcome with the weighted-average correlation
test, if a weight function $g\rund{\f}$ like (\ref{eq650}) is
applied. On the angular scale $\f\le\f_{\rm o}=0.24$ deg the strong
dependence of the expected correlations on $\f$ enables the
weighted average to detect a gradient of the galaxy 
number density. For angular distances $\f\gg\f_{\rm o}$ the weight
function $g\rund{\f}$ is flat. Therefore, in the outer regions not the
exact distribution of galaxies but only their mean number density is
relevant. This means the weighted average both is sensitive to a
quasar-galaxy correlation on a fixed angular radius $\f_{\rm o}$ and
at the same time includes the information on the mean galaxy density
from the outer parts of the field. Consequently, and in contrast to
Spearman's rank test, our method can be used to effectively analyse
large galaxy fields.

\begin{figure*}
   \centerline{\psfig{figure=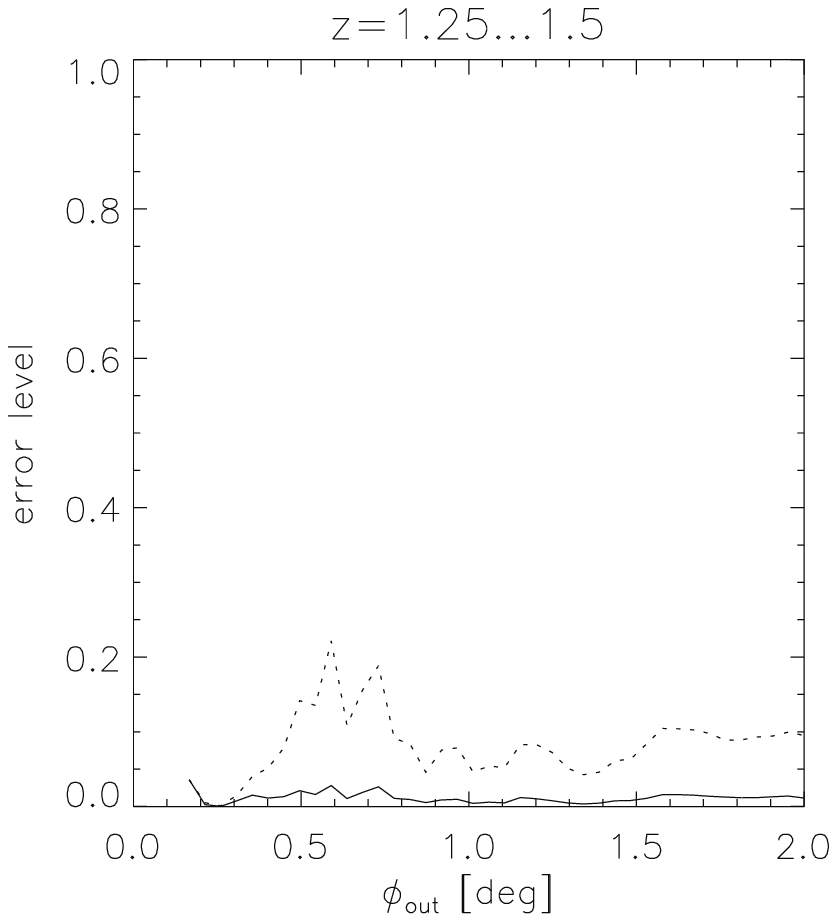,height=7.0cm,width=7.0cm}
   \hfill\psfig{figure=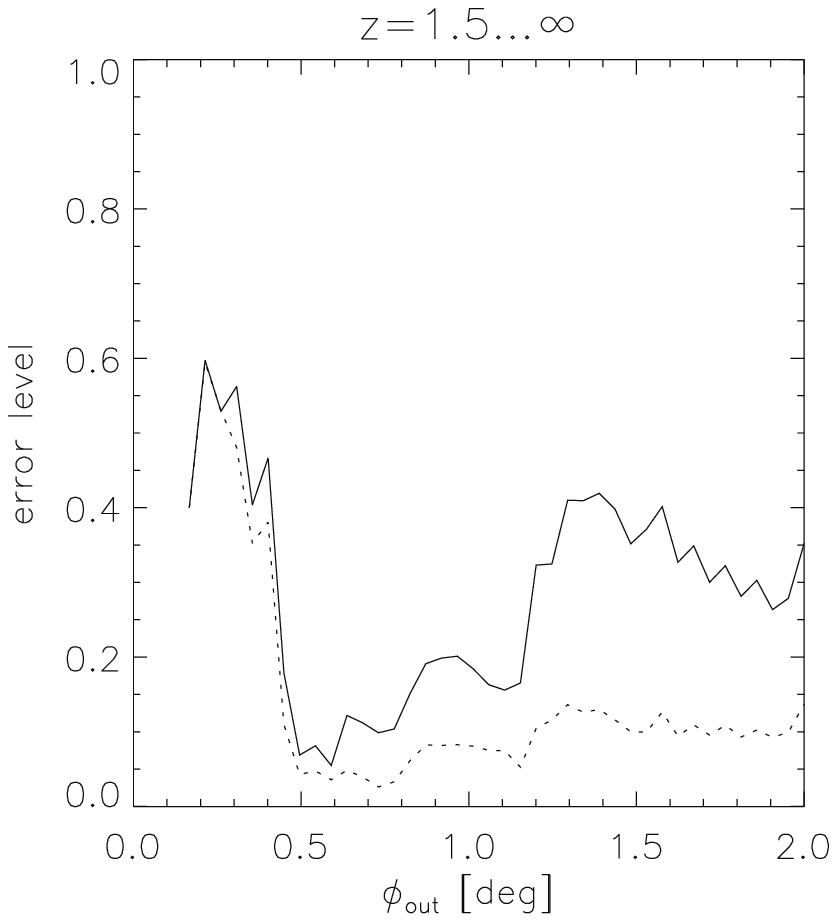,height=7.0cm,width=7.0cm}\hfill}
   \caption[]{Error level for rejecting the null-hypothesis of a
      Poissonian galaxy distribution as a function of the outer field
      radius $\f_{\rm out}$ for four different quasar subsamples.
      The redshift range of the quasars is denoted by $z$, their maximum
      visual magnitude is $m_{\rm max}=21$ for the solid lines and
      $m_{\rm max}=19$ for the dotted lines.}
   \label{fig6}
\end{figure*}

For four different quasar subsamples Fig.\ \ref{fig6} displays the
error level for rejecting the null-hypothesis of a Poissonian galaxy
distribution as a function of the outer field radius $\f_{\rm out}\le
2$ deg. The curves illustrate that a relatively small variation  
of $\f_{\rm out}$ can have a significant effect on the resulting
value of the error level, even if $\f_{\rm out}$ is beyond 1 deg.
Tables~\ref{ta3} and \ref{ta4} summarize the results obtained by
reanalysing all the quasar subsamples as listed in 
tables~\ref{ta1} and \ref{ta2} with an outer radius of 2 deg.

%
%
%
%

%
%
\begin{table}
   \caption[]
      {As table \protect\ref{ta1} but with an outer galaxy field
      radius of $\f_{\rm out} = 2^{\rm o}$.}
   \label{ta3} 
   \begin{center}
      \begin{tabular}{|rr|r|r|r|}
         \hline
          & & & \multicolumn{2}{c|}{$\f_{\rm out}=2^{\rm o}$}\\
         \cline{4-5}
         $z_{\rm min}$ & $m_{\rm max}$ & $N_{\rm Q}$ & $N$ & 
         $\e_{\rm wa}$ \\
         \hline\hline
         0.50 & 21.00 & 238 & 2818 &  7 \\
         0.50 & 20.00 & 218 & 2585 &  2 \\
         0.50 & 19.00 & 192 & 2278 &  1 \\
         0.50 & 18.75 & 169 & 1984 &  1 \\
         0.50 & 18.50 & 161 & 1825 &  2 \\
         0.50 & 18.25 & 123 & 1426 &  8 \\
         0.50 & 18.00 & 114 & 1326 & 15 \\
         \hline		                  
         0.75 & 21.00 & 179 & 2127 & 21 \\
         0.75 & 20.00 & 166 & 1977 & 12 \\
         0.75 & 19.00 & 144 & 1727 &  8 \\
         0.75 & 18.75 & 126 & 1482 &  8 \\
         0.75 & 18.50 & 120 & 1356 & 10 \\
         0.75 & 18.25 &  87 & 1002 & 23 \\
         0.75 & 18.00 &  80 &  913 & 37 \\
         \hline		                  
         1.00 & 21.00 & 130 & 1555 & 19 \\
         1.00 & 20.00 & 123 & 1482 & 15 \\
         1.00 & 19.00 & 107 & 1312 & 15 \\
         1.00 & 18.75 &  94 & 1140 &  7 \\
         1.00 & 18.50 &  88 & 1014 &  9 \\
         1.00 & 18.25 &  60 &  703 & 19 \\
         1.00 & 18.00 &  54 &  626 & 24 \\
         \hline		                  
         1.25 & 21.00 &  97 & 1082 &  4 \\
         1.25 & 20.00 &  93 & 1035 &  3 \\
         1.25 & 19.00 &  80 &  909 &  5 \\
         1.25 & 18.75 &  68 &  745 &  2 \\
         1.25 & 18.50 &  64 &  699 &  1 \\
         1.25 & 18.25 &  42 &  481 &  9 \\
         1.25 & 18.00 &  37 &  417 & 14 \\
         \hline		                  
         1.50 & 21.00 &  59 &  625 & 35 \\
         1.50 & 20.00 &  56 &  599 & 30 \\
         1.50 & 19.00 &  46 &  507 & 14 \\
         1.50 & 18.75 &  36 &  368 &  5 \\
         1.50 & 18.50 &  34 &  334 &  3 \\
         1.50 & 18.25 &  20 &  203 &  7 \\
         1.50 & 18.00 &  18 &  189 &  4 \\
         \hline
      \end{tabular}
   \end{center}
\end{table}
%
%

%
%
\begin{table}
   \caption[]
      {As table \protect\ref{ta2} but with an outer galaxy field
      radius of $\f_{\rm out} = 2^{\rm o}$.}
   \label{ta4}  
   \begin{center}
      \begin{tabular}{|rrr|r|r|r|}
         \hline
          & & & & \multicolumn{2}{c|}{$\f_{\rm out}=2^{\rm o}$}\\
         \cline{5-6}
         $z_{\rm min}$ & $z_{\rm max}$ & $m_{\rm max}$ & 
         $N_{\rm Q}$ & $N$ & $\e_{\rm wa}$ \\
         \hline\hline
         0.50 & 0.75 & 21.00 & 61 & 713 &  7 \\
         0.50 & 0.75 & 20.00 & 53 & 617 &  2 \\
         0.50 & 0.75 & 19.00 & 49 & 560 &  2 \\
         0.50 & 0.75 & 18.75 & 44 & 511 &  1 \\
         0.50 & 0.75 & 18.50 & 42 & 478 &  3 \\
         0.50 & 0.75 & 18.25 & 37 & 433 &  9 \\
         0.50 & 0.75 & 18.00 & 35 & 422 & 11 \\
         \hline 	                        
         0.75 & 1.00 & 21.00 & 49 & 572 & 44 \\
         0.75 & 1.00 & 20.00 & 43 & 495 & 27 \\
         0.75 & 1.00 & 19.00 & 37 & 415 & 14 \\
         0.75 & 1.00 & 18.75 & 32 & 342 & 35 \\
         0.75 & 1.00 & 18.50 & 32 & 342 & 35 \\
         0.75 & 1.00 & 18.25 & 27 & 299 & 48 \\
         0.75 & 1.00 & 18.00 & 26 & 287 & 67 \\
         \hline 	                         
         1.00 & 1.25 & 21.00 & 34 & 483 & 92 \\
         1.00 & 1.25 & 20.00 & 31 & 457 & 89 \\
         1.00 & 1.25 & 19.00 & 28 & 413 & 78 \\
         1.00 & 1.25 & 18.75 & 27 & 405 & 75 \\
         1.00 & 1.25 & 18.50 & 25 & 325 & 96 \\
         1.00 & 1.25 & 18.25 & 18 & 222 & 67 \\
         1.00 & 1.25 & 18.00 & 17 & 209 & 60 \\
         \hline 	                         
         1.25 & 1.50 & 21.00 & 38 & 457 &  1 \\
         1.25 & 1.50 & 20.00 & 37 & 436 &  1 \\
         1.25 & 1.50 & 19.00 & 34 & 402 &  9 \\
         1.25 & 1.50 & 18.75 & 32 & 377 &  6 \\
         1.25 & 1.50 & 18.50 & 30 & 365 &  5 \\
         1.25 & 1.50 & 18.25 & 22 & 278 & 30 \\
         1.25 & 1.50 & 18.00 & 19 & 228 & 59 \\
         \hline 	                              
         1.50 &$\infty$& 21.00 & 59 & 625 & 35 \\
         1.50 &$\infty$& 20.00 & 56 & 599 & 30 \\
         1.50 &$\infty$& 19.00 & 46 & 507 & 14 \\
         1.50 &$\infty$& 18.75 & 36 & 368 &  5 \\
         1.50 &$\infty$& 18.50 & 34 & 334 &  3 \\
         1.50 &$\infty$& 18.25 & 20 & 203 &  7 \\
         1.50 &$\infty$& 18.00 & 18 & 189 &  4 \\
         \hline
      \end{tabular}
   \end{center}
\end{table} 
%
%

\begin{figure*}
   \centerline{\psfig{figure=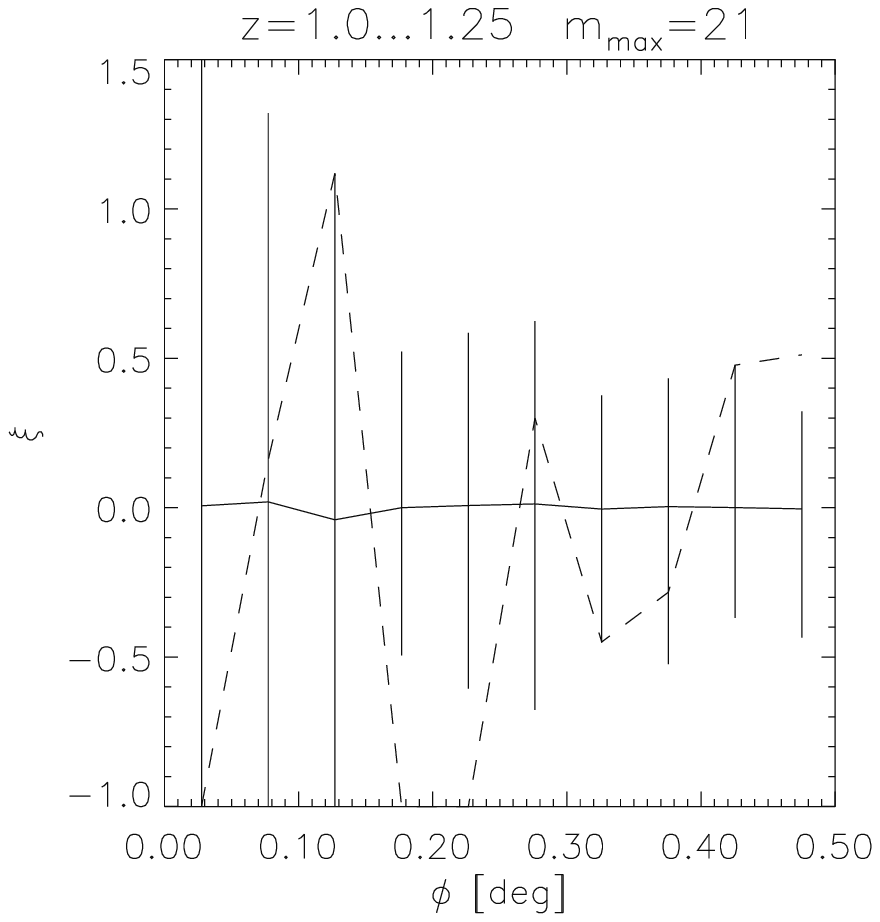,height=7.0cm,width=7.0cm}
   \hfill\psfig{figure=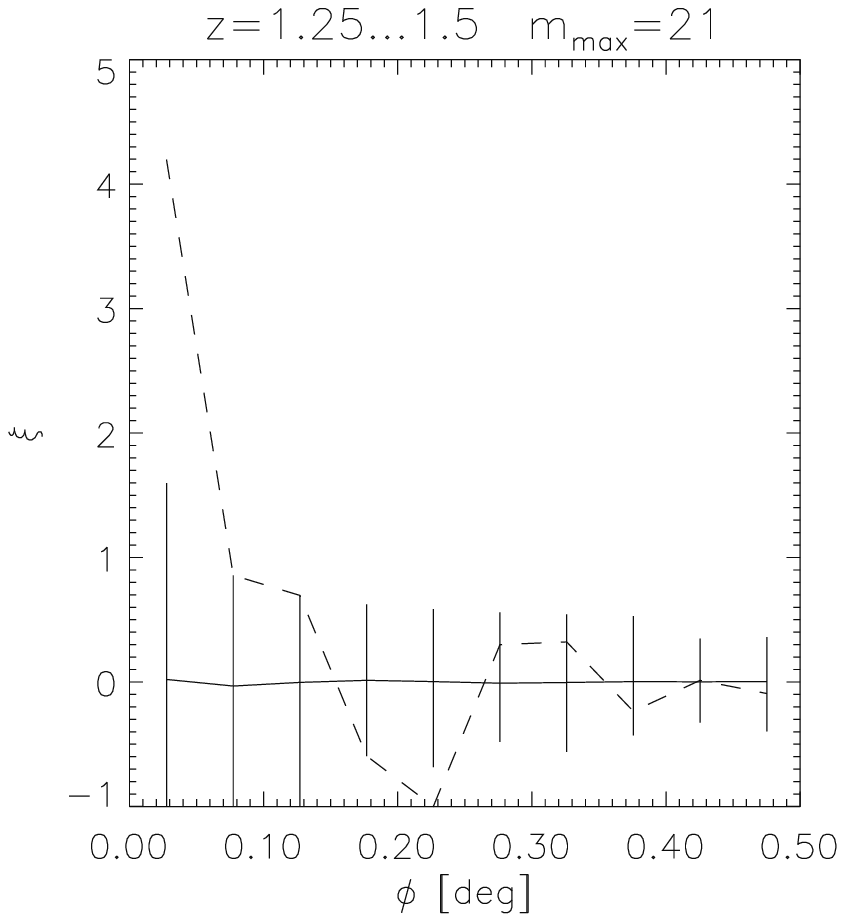,height=7.0cm,width=7.0cm}\hfill}
   \centerline{\psfig{figure=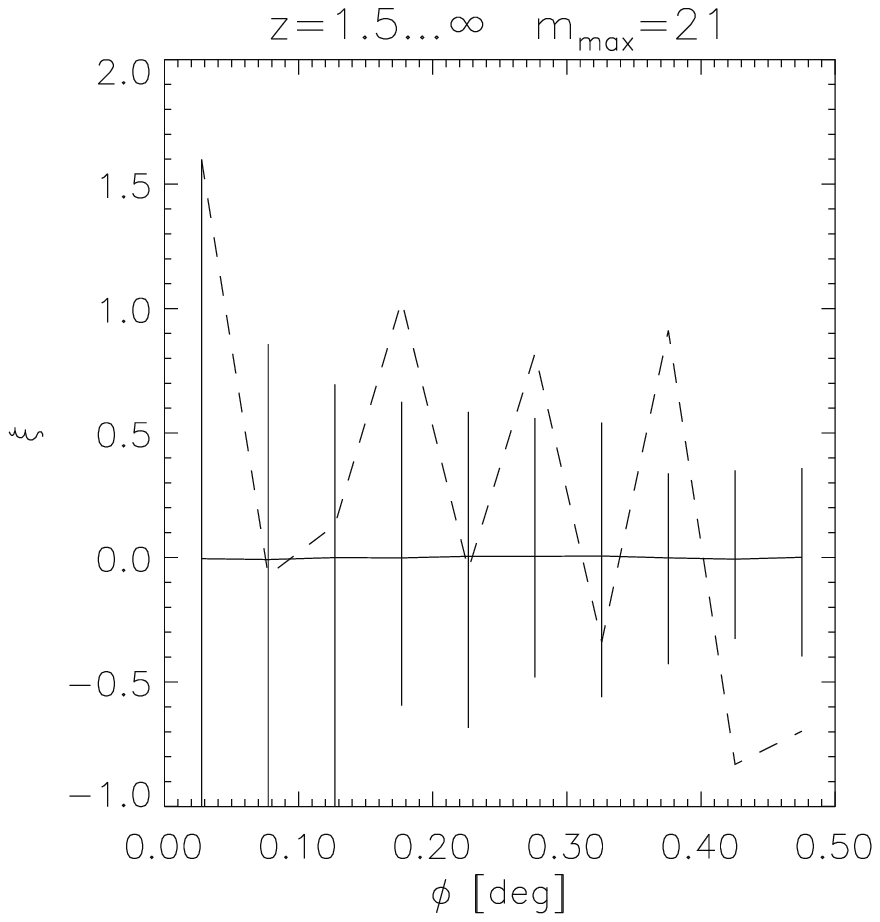,height=7.0cm,width=7.0cm}
   \hfill\psfig{figure=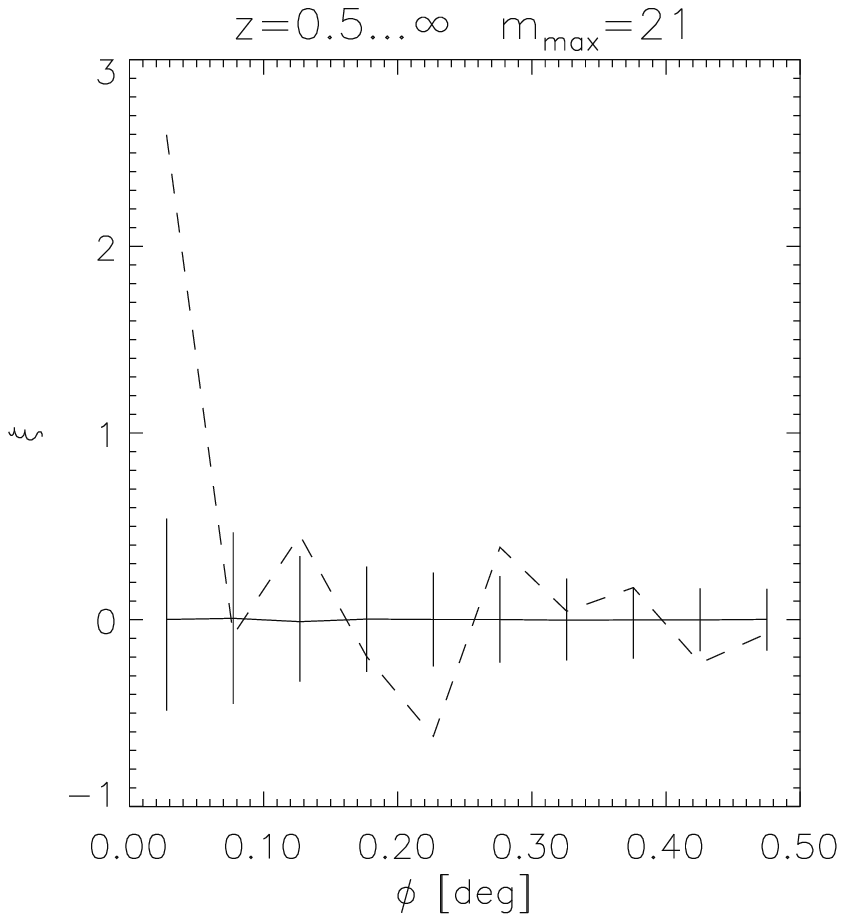,height=7.0cm,width=7.0cm}\hfill}
   \caption[]{Quasar-galaxy correlation function for 1Jy quasars and
      IRAS galaxies determined from different quasar subsamples.
      The redshift range of the quasars is denoted by $z$, their maximum
      visual magnitude by $m_{\rm max}$.}
   \label{fig6.5}
\end{figure*}

A more detailed view of the angular galaxy distribution
at small distances to 1Jy quasars can be obtained
from the correlation functions plotted in Fig.\ \ref{fig6.5}. They are
calculated using ring-like total galaxy fields of radii 
$\f_{\rm in}=10$ arcsec and $\f_{\rm out}=0.5$ deg constructed
according to Sec.\ \ref{qgcorr} by merging the individual galaxy
fields around the quasars of a given subsample. As before, the 
inner part is removed to ensure that the fields are not contaminated
by the quasars themselves. A total galaxy field is then divided into
$n=10$ sub-rings 
$\eck{\r_i,\r_{i+1}}$ given by Eq.\ (\ref{eq180}). Denoting the 
number of galaxies within sub-ring no.\ $i$ by $z_i$ and its solid
angle by $A_i:=\pi\rund{\r_{i+1}^2-\r_i^2}$ we  
derive the value $\xi_i$ from
\bdm
   \xi_i:=\frac{z_i\,\sum_j A_j}{A_i\,\sum_j z_j}-1
\edm
and assign it to the ``mean radius'' $\rund{\r_i+\r_{i+1}}/2$ of the
ring. 

%
%
%
%

In the diagrams of Fig.\ \ref{fig6.5} the corresponding points are
connected by a dashed line which can be interpreted as an
approximation to the quasar-galaxy correlation function. 
To give an estimate of the errors we do not attach
error bars of length $\sqrt{z_i}$ to a graph as it is commonly
done, because their significance level is strongly dependent on $z_i$ 
if $z_i$ is small. Instead we numerically generate a set of 1000
artificial galaxy fields with a random (Poissonian) galaxy
distribution. The number
of galaxies on each of them is equal to that of the total galaxy field
of the regarded quasar subsample. They are subjected to the same
procedure as the real fields and for each sub-ring we derive
the mean value of $\xi_i$ and a $70\%$ error bar: $15\%$
of the simulations yield a value of $\xi_i$ below the lower end of the
error bar, the values of an equal fraction are above the upper end.     

The selected quasar subsamples are characterised in the diagram
headings by the redshift $z$ of the quasars and their apparent visual
magnitude $m$. The bottom right panel which is calculated from our
largest subsample shows a strong signal of a correlation. However,
if compared to Fig.\ \ref{fig2} the length scale of the correlation 
appears to be smaller and its amplitude to be much greater than
expected from 
the gravitational lens effect. This is not too surprising, because as
it was mentioned above there might be a spatial association between
IRAS galaxies and low redshift quasars. The remaining three panels are 
obtained from three neighbouring high-redshift
subsamples. They visualize the results from Table~\ref{ta2}: 
quasar-galaxy anticorrelation for the quasar redshift range $1.0\le z <
1.25$, correlations for $1.25\le z <1.5$ and $z\ge 1.5$. From the top
right diagram it is evident that the correlation for quasars with
$1.25\le z <1.5$ is produced by a significant galaxy overdensity close
to the quasars, again with an amplitude greater than expected from the
numerical simulations reported by Bartelmann (\cite{msb95}), whereas
it seems to be caused by a sharp decrease of the galaxy number density
at greater distances in the highest redshift subsample. This decrease
corresponds to the ``hole'' 
in the galaxy distribution which we have already seen in
Fig.\ \ref{fig5.5}. 
\subsection{Additional simulations}
Given our null-hypothesis of a purely random distribution of galaxies
around the 1Jy quasars we have so far quantified our correlation
results in terms of the error level. The error level assigned to a
value $R$ of the correlation coefficient $r$ has been introduced to be
the probability $\e:=P_{\rm o}\rund{r\ge R}$ for $r$ to take a value
equal to or greater than $R$ for a Poissonian galaxy distribution.  

Suppose now we reject the null hypothesis for some of the
quasar subsamples, because the associated error level is low. 
Assuming the quasar-galaxy correlation to be induced by gravitational
lensing we would expect it to be described approximately by a
correlation function as derived by Bartelmann 
(see Sec.\ \ref{qgcorr}), a fit to which we gave in 
Eq.\ (\ref{eq600}). With this new premise we can then again ask the
question: What is the probability $P_{\rm a}\rund{r\ge R}$ for the
correlation coefficient to take a value equal to or greater than $R$
or, equivalently, what is the probability to find an error level equal
to or lower than $\e$.

\begin{figure*}
   \centerline{\psfig{figure=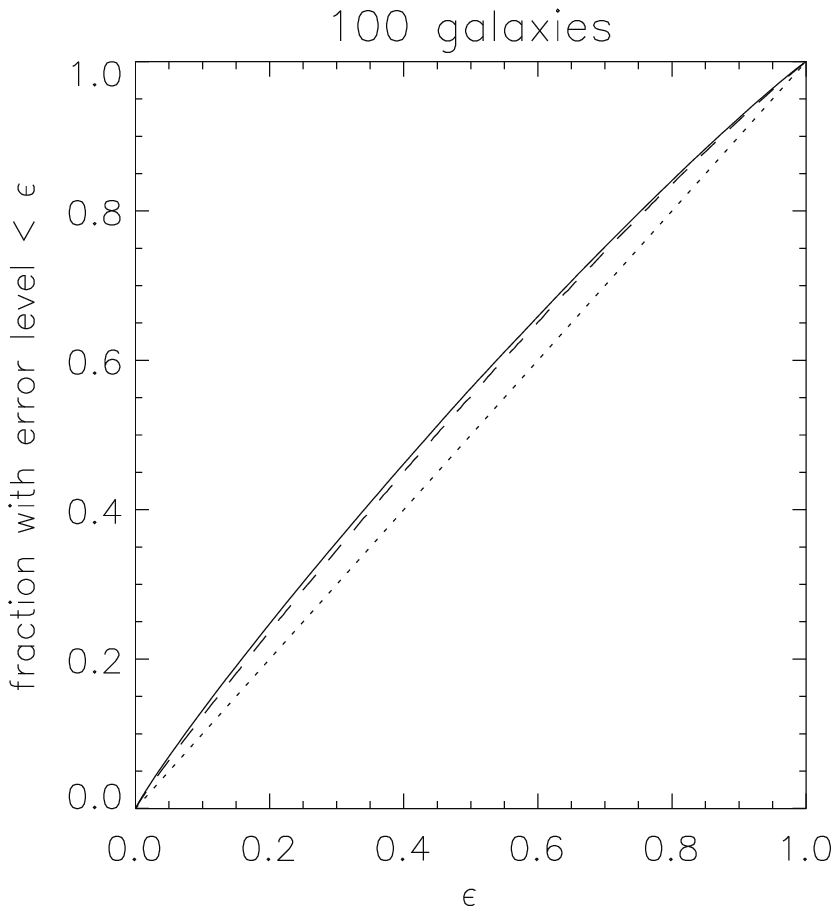,height=7.0cm,width=7.0cm}
   \hfill\psfig{figure=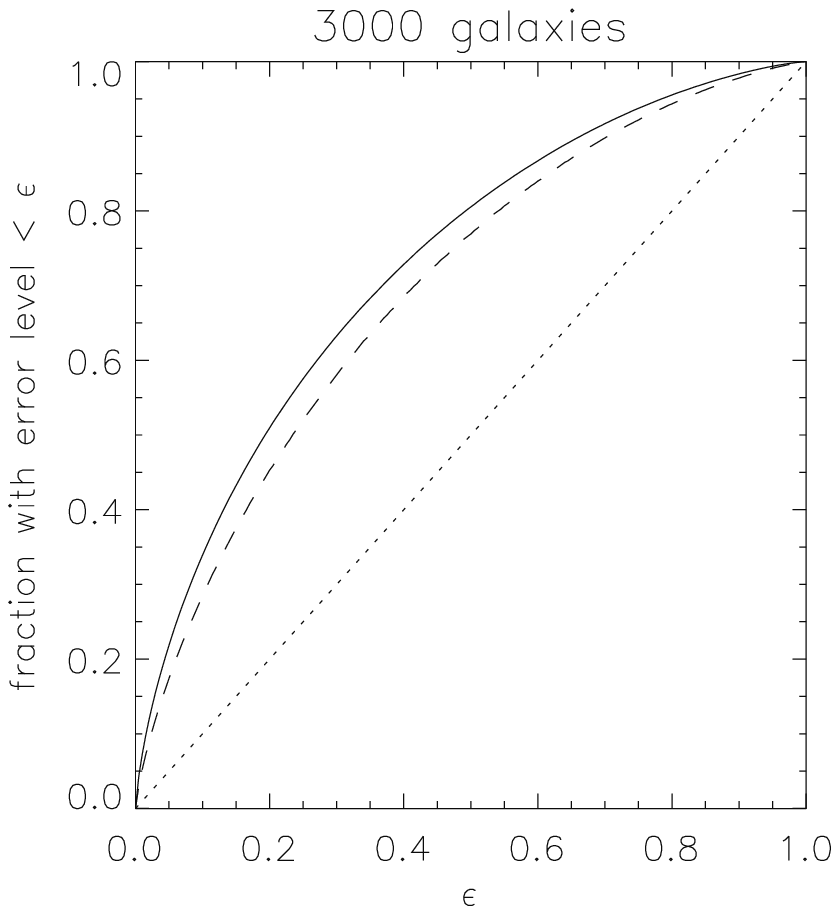,height=7.0cm,width=7.0cm}\hfill}
   \caption[]{Fraction of simulated galaxy fields yielding an error
   level equal to or lower than some limit $\e$ for a galaxy
   distribution according to the theoretically expected quasar-galaxy
   correlation analysed with the weighted-average test (solid line)
   or Spearman's rank-order test (dashed line) and for a purely random
   galaxy distribution analysed with either one of the tests (dotted
   line). The number of galaxies per simulated field is 100 for the
   left panel and 3000 for the right panel.} 
   \label{fig7}
\end{figure*}

In order to find an answer we generated a large number of synthetic
galaxy fields of radius $\f_{\rm out}=30'$ following the correlation
function (\ref{eq600}) and containing 100 galaxies each. We subjected
them to both Spearman's rank-order test and the weighted-average
correlation test with the optimum weight function (\ref{eq700})
($h=1$). The fraction of simulated fields resulting in an error level
equal to or lower than $\e$ then directly yields the required
probability, which is plotted in Fig.\ \ref{fig7}, left panel, as a
function of $\e$. The solid line presents the result for the
weighted-average test, the dashed line for Spearman's rank test. The
dotted 
line would arise for both correlation tests if the simulated galaxies
were distributed randomly; it just reflects the definition of the
error level. Repeating the whole procedure with 3000 galaxies per
field produces the right panel of Fig.\ \ref{fig7}.

%
%
%
%

As to quasar subsamples with a total number of galaxies $N\approx 100$
on fields of $\f_{\rm out}\approx 30'$ Fig.\ \ref{fig7} clearly shows
that it is hardly more probable to find low values of the error level
in the case of the expected, lensing-induced quasar-galaxy correlation
than it is for randomly distributed galaxies. When interpreting this
result one should be aware of two facts: On the one hand, $N=100$ is a
typical number for rather large quasar subsamples (cf.\ Tables
\ref{ta1} and \ref{ta2}) and for subsamples 
with fewer galaxies the difference between the correlated and the 
random distributions will be even smaller. On the other hand, the
correlation function (\ref{eq600}) is a fit to what Bartelmann derived
from numerical simulations for an artificial galaxy catalog, which
includes all galaxies with a visual magnitude $m\le 21$. Therefore it
is quite probable that the IRAS catalog is deeper in redshift than
this synthetic catalog and shows a stronger correlation with quasars.
Nonetheless, as the curves in Fig.\ \ref{fig7} are so extremely close
together, we conclude that either 
\begin{itemize}
   \item the low error levels are the result of a statistical
   fluctuation and do not correspond to a real correlation, or 
   \item the numerical results in Bartelmann (\cite{msb95})
   significantly underestimate the expected correlations between QSOs
   and IRAS galaxies, either because the non-linear density
   inhomogeneities are not properly resolved in these simulations, or
   because the assumed redshift distribution is much shallower than
   that of IRAS galaxies, or
   \item the expected angular scale $\f_{\rm o}$ is smaller than
   obtained by Bartelmann, as might be indicated by the correlation
   function plotted in Fig.\ \ref{fig6.5}, or 
   \item for some reason the correlation is not described correctly by
   assuming a gravitational lensing effect by the large-scale
   structure. However, in view of the results by Fort et
   al.\ (\cite{for95}) 
   quoted in the introduction, we consider this latter possibility
   unlikely.
\end{itemize} 
%
%
%
\section{Summary and conclusions}
\label{summary}
In this paper, we introduced a new statistical method, the
weighted-average correlation test, to look for angular correlations
between
background quasars and foreground galaxies. As was pointed out by
Bartelmann \& Schneider, such correlations are predicted by
gravitational lens theory under plausible assumptions concerning the
quasar luminosity function and the dependence between the
distributions of dark and luminous matter. 

Whereas the methods applied in earlier investigations of this
phenomenon need to group the galaxies according to their positions
into some arbitrarily defined bins, the weighted average uses the
exact distance of each individual galaxy to the corresponding
quasar. Furthermore, if additional information about the assumed
correlation is available it can be included into our test via choosing
a proper weight function. In an analytic calculation we could derive a
formula which allows to construct an optimum weight function from the
expected quasar-galaxy correlation function. The verification of this
result by means of numerical simulations has also shown that the
weighted-average correlation test can be more significant than
Spearman's rank-order test, even if the weight function deviates
considerably from its optimum. 

To look for a quasar-galaxy correlation we analysed the distribution
of IRAS galaxies on circular fields around 1Jy quasars. This was
carried out for different quasar subsamples by merging the individual
galaxy fields and subjecting the resulting total field to the
weighted-average test. As we have explained and illustrated, the tests
should 
be performed on galaxy fields as large as possible. Actually, the
outer regions carry useful information about the mean galaxy number
density, which is obviously needed to detect a possible galaxy
overdensity in the center of the fields. Supplied with an appropriate
weight function the weighted average can make use of this information.

With small galaxy fields of radius $\f_{\rm out}\approx 0.5$ deg as
well as with large ones of radius $\f_{\rm out}=2$ deg many of the
quasar subsamples give rise to a low error level for the rejection of
a purely random galaxy distribution. This is in agreement with 
previous findings in BS. As argued there, a correlation of galaxies
with low-redshift 
quasars might be due to a spatial association, whereas a correlation
with high-redshift quasars could be caused by gravitational lensing.
Nevertheless there seems to be an anticorrelation between galaxies and
quasars of redshift $1.0\le z<1.25$, and its interpretation in terms of
these hypotheses is not obvious. However, it should be pointed out
that this anticorrelation occurs with an error level larger than 10\%
and is therefore not of high significance.
To visualize the angular galaxy
distribution within different subsamples we compiled plots of the
quasar-galaxy correlation function. Although they look quite
different, this is, of course, not a statistically significant
indication that they do not represent realisations of the same parent
distribution.

From additional simulations we expected to get further hints for the
interpretation of our results. We numerically generated a large number
of synthetic galaxy fields with galaxies distributed according to the 
correlation function expected from gravitational lensing by the
large-scale structure as studied in Bartelmann (\cite{msb95}).
On the basis of these results, we found that the lensing-induced
correlations between 1Jy quasars and IRAS galaxies should be
detectable neither with the weighted-average test nor with Spearman's
rank-order test, because of the sparseness of the samples. 

Therefore, as stated above, we conclude that either 
\begin{itemize}
   \item the low error levels are the result of a statistical
   fluctuation and do not correspond to a real correlation, or 
   \item the numerical results in Bartelmann (\cite{msb95})
   significantly underestimate the expected correlations between QSOs
   and IRAS galaxies, either because the non-linear density
   inhomogeneities are not properly resolved in these simulations, or
   because the assumed redshift distribution is much shallower than
   that of IRAS galaxies, or
   \item the expected angular scale $\f_{\rm o}$ is smaller than
   obtained by Bartelmann, as might be indicated by the correlation
   function plotted in Fig.\ \ref{fig6.5}, or 
   \item for some reason the correlation is not described correctly by
   assuming a gravitational lensing effect by the large-scale
   structure. However, in view of the results by Fort et 
   al.\ (\cite{for95})
   quoted in the introduction, we consider this latter possibility
   unlikely.
\end{itemize}
%
%
%
{\bf Acknowledgements.} \SFB
%
%
%
%
\appendix
%
%
\section*{Appendix A. Optimizing the weight function}
In Sect.\ \ref{defs} it was argued that the correlation
coefficient $r$ as defined in (\ref{eq100}) can be optimized for
distinguishing between two statistical hypotheses via maximizing
$\abs{Q_1}$, given by Eq.\ (\ref{eq150}). Here we intend to carry out
the maximisation for
\beq\label{eqa10}
   r_g\rund{\f_1,\ldots,\f_N} := 
   \frac{1}{N}\sum_{j=1}^N g\rund{\f_j}\ \ ,
\eeq
which was shown in Sect.\ \ref{qgcorr} to be a limiting case of
$r$ applied to the analysis of quasar-galaxy correlations.
Accordingly, the symbols $\f_1,\ldots,\f_N$ denote galaxy positions.

In addition, it will be demonstrated that the weight function
$g\rund{\f}$ we find in this way also represents nearly a stationary
point of the mean error level (\ref{eq140}), provided the two
hypotheses are ``similar''.

We assume the galaxy positions to be independent of one another, so
their distribution can be characterised by a one-dimensional radial
probability density $p\rund{\f}$. Let $p_{\rm o}\rund{\f}$
describe the radial galaxy distribution of the null-hypothesis,
i.e. without any quasar-galaxy association. Furthermore, suppose we
have some hints, e.g.\ from theory, that the real distribution of
galaxies might be represented by the different probability density
$p_{\rm a}\rund{\f}$. Following Sect.\ \ref{qgcorr} we introduce a
geometrical factor $G\rund{\f}\ge 0$ and write
\beao
   \begin{array}{rcll}
      p_{\rm o}\rund{\f} &=& c_{\rm o}\cdot G\rund{\f}\ \ ,
      &\displaystyle\int_{\f_{\rm in}}^{\f_{\rm out}}
      c_{\rm o}\,G\rund{\f}\de\f=1\ \ ,\vspace{1\jot}\\
      p_{\rm a}\rund{\f} &=& c_{\rm a}\rund{\f}\cdot G\rund{\f}\ \ , 
      &\displaystyle\int_{\f_{\rm in}}^{\f_{\rm out}}
      c_{\rm a}\rund{\f}G\rund{\f}\de\f=1\ \ .
   \end{array}
\eeao
\subsection*{A.1\ \ The quantity $Q_1$} 
Using the above expressions we have
\beao
   \ave{r_g}_{\rm o}  & = & \int_{\f_{\rm in}}^{\f_{\rm out}}
   \!\cdots\!\int_{\f_{\rm in}}^{\f_{\rm out}}
   \eck{\frac{1}{N}\sum_{i=1}^N g\rund{\f_i}}\prod_{j=1}^N
   \eck{c_{\rm o}\,G\rund{\f_j} \de\f_j} = 
   c_{\rm o}\,\int_{\f_{\rm in}}^{\f_{\rm out}}
   g\rund{\f} G\rund{\f} \de\f\ \ ,\\
   \ave{r_g}_{\rm a}  & = & \int_{\f_{\rm in}}^{\f_{\rm out}}
   g\rund{\f} c_{\rm a}\rund{\f} G\rund{\f} \de\f\ \ ,\\
   \sigma_{\rm o}^2 & = & \int_{\f_{\rm in}}^{\f_{\rm out}}
   \!\cdots\!\int_{\f_{\rm in}}^{\f_{\rm out}}
   \eck{r_g\rund{\f_1,\ldots,\f_N}-\ave{r_g}_{\rm o}}^2
   \prod_{j=1}^N \eck{c_{\rm o}\,G\rund{\f_j} \de\f_j} =\\
    & = & \int_{\f_{\rm in}}^{\f_{\rm out}}
   \!\cdots\!\int_{\f_{\rm in}}^{\f_{\rm out}}
   \eck{\frac{1}{N}\sum_{i=1}^N g\rund{\f_i}}^2
   \prod_{j=1}^N \eck{c_{\rm o}\,G\rund{\f_j} \de\f_j} - 
   \ave{r_g}_{\rm o}^2=\\
    & = & \frac{1}{N^2}\sum_{i=1}^N \int_{\f_{\rm in}}^{\f_{\rm out}} 
   \eck{g\rund{\f_i}}^2  c_{\rm o}\,G\rund{\f_i} \de\f_i +\\
    &   & +\frac{1}{N^2}
   \sum_{{\scriptstyle i,j=1}\atop {\scriptstyle i \ne j}}^N
   \int_{\f_{\rm in}}^{\f_{\rm out}}\!\!
   \int_{\f_{\rm in}}^{\f_{\rm out}}
   g\rund{\f_i} g\rund{\f_j} 
   c_{\rm o}^2 G\rund{\f_i} G\rund{\f_j}
   \de\f_i\,\de\f_j - \ave{r_g}_{\rm o}^2 =\\
    & = & \frac{1}{N} \int_{\f_{\rm in}}^{\f_{\rm out}}
   \eck{g\rund{\f}}^2  c_{\rm o}\,G\rund{\f} \de\f -
   \frac{1}{N}\,\ave{r_g}_{\rm o}^2\ \ .
\eeao
The substitution of $r$ by $r_g$ in definition (\ref{eq150}) then
yields
\bea
   Q_1  & = & \sqrt{N}\frac{\int g\rund{\f} c_{\rm a}\rund{\f}
   G\rund{\f}\de\f - 
   \int g\rund{\f} c_{\rm o} G\rund{\f} \de\f}
   {\sqrt{\int\eck{g\rund{\f}}^2 c_{\rm o} G\rund{\f} \de\f -
   \eck{\int g\rund{\f} c_{\rm o} G\rund{\f} \de\f}^2}}=
   \nonumber\\
    & = & \int\frac{g\rund{\f'}-\int g\rund{\f} 
   c_{\rm o} G\rund{\f} \de\f}
   {\sqrt{\int\eck{g\rund{\f}}^2 c_{\rm o} G\rund{\f} \de\f -
   \eck{\int g\rund{\f} c_{\rm o} G\rund{\f}\de\f}^2}}
   \sqrt{N}\,c_{\rm a}\rund{\f'} 
   G\rund{\f'} \de\f' =\nonumber\\
    & = & \int L\rund{\f,E,W;g\rund{\f}} \de\f\ \ ,\label{eqa25}
\eea
where all the integrations are to be performed over the interval 
$\eck{\f_{\rm in},\f_{\rm out}}$ and we have introduced the
abbreviations 
\bdm
   L\rund{\f,E,W;g\rund{\f}} := \sqrt{N}\,c_{\rm a}\rund{\f}  
   G\rund{\f}\,\frac{g\rund{\f}-E}{\sqrt{W-E^2}}\ \ ,
\edm
\vspace{-\belowdisplayskip}
\vspace{-\abovedisplayskip}
\beao
   E & := & \int_{\f_{\rm in}}^{\f_{\rm out}} 
   g\rund{\f} c_{\rm o}\,G\rund{\f} \de\f\ \ ,\\
   W & := &  \int_{\f_{\rm in}}^{\f_{\rm out}} 
   \eck{g\rund{\f}}^2 c_{\rm o}\,G\rund{\f} \de\f\ \ .   
\eeao

In order to maximize $\abs{Q_1}$ we investigate the influence
of a variation $\e\cdot h\rund{\f}$ of $g\rund{\f}$. For that purpose,
we substitute 
$g\rund{\f}\rightarrow\g\rund{\f,\e}:=g\rund{\f}+\e\cdot h\rund{\f}$
and find 
\bdm
   Q_1\rund{\e} = \int _{\f_{\rm in}}^{\f_{\rm out}}
   L\RUND{\f,E\rund{\e},W\rund{\e};\g\rund{\f,\e}} \de\f 
\edm
with
\beao
   E\rund{\e} & := & \int_{\f_{\rm in}}^{\f_{\rm out}}
   \g\rund{\f,\e} c_{\rm o}\,G\rund{\f} \de\f\ \ ,\\
   W\rund{\e} & := & \int_{\f_{\rm in}}^{\f_{\rm out}}
   \eck{\g\rund{\f,\e}}^2 c_{\rm o}\,G\rund{\f} \de\f\ \ .
\eeao
The condition for the weight function $g\rund{\f}$ to produce an
extreme value of $Q_1$ can now be expressed as
\beq\label{eqa50}
   \at{\tot{}{\e}Q_1\rund{\e}}_{\e=0} = 
   \int_{\f_{\rm in}}^{\f_{\rm out}}
   \at{\tot{}{\e}L\RUND{\f,E\rund{\e},W\rund{\e};\g\rund{\f,\e}}}
   _{\e=0} \de\f = 0\ \ ,
\eeq
which is equivalent to
\bea
   \lefteqn{\int\at{\prt{L}{\g}\rund{\f}}_{\e=0} 
   h\rund{\f} \de\f\  
    + \int\!\!\int\at{\prt{L}{E}\rund{\f}}_{\e=0}
   h\rund{t} c_{\rm o}\,G\rund{t} \de t \de\f\ +}\nonumber\\
    & & +\ \int\!\!\int\at{\prt{L}{W}\rund{\f}}_{\e=0} 2\, 
   g\rund{t} h\rund{t} c_{\rm o}\,G\rund{t} \de t \de\f =0\ \ ,
   \label{eqa100}
\eea
because of the relations
\bdm
   \tot{L}{\e} = \prt{L}{\g}\tot{\g\rund{\f,\e}}{\e} +
   \prt{L}{E}\tot{E\rund{\e}}{\e}+
   \prt{L}{W}\tot{W\rund{\e}}{\e}\ \ ,\\
\edm
and
\beao
   \at{\tot{\g\rund{\f,\e}}{\e}}_{\e=0} & = & h\rund{\f}\ \ ,\\
   \at{\tot{E\rund{\e}}{\e}}_{\e=0} & = & 
   \int_{\f_{\rm in}}^{\f_{\rm out}} h\rund{\f} c_{\rm o}\,
   G\rund{\f} \de\f\ \ ,\\
   \at{\tot{W\rund{\e}}{\e}}_{\e=0} & = & 2\cdot 
   \int_{\f_{\rm in}}^{\f_{\rm out}} g\rund{\f}  
   h\rund{\f} c_{\rm o}\,G\rund{\f} \de\f\ \ .
\eeao
Exchanging both the order of integrations and the 
variables $\f$ and $t$ in the second and third term of Eq.\
(\ref{eqa100}) leads to
\bea
   \lefteqn{\int\at{\prt{L}{\g}\rund{\f}}_{\e=0} 
   h\rund{\f} \de\f\ + 
   \int\!\!\int\at{\prt{L}{E}\rund{t}}_{\e=0}
   h\rund{\f} c_{\rm o}\,G\rund{\f} \de t \de\f\ +}\nonumber\\
    & & +\ \int\!\!\int\at{\prt{L}{W}\rund{t}}_{\e=0} 2 
   g\rund{\f} h\rund{\f} c_{\rm o}\,G\rund{\f} 
   \de t \de\f =
   \int_{\f_{\rm in}}^{\f_{\rm out}}
   q\rund{\f} h\rund{\f} \de\f = 0\ \ ,\label{eqa200}
\eea
where $q\rund{\f}$ is defined to be
\bdm
   q\rund{\f} := \at{\prt{L}{\g}\rund{\f}}_{\e=0}+c_{\rm o}\,
   G\rund{\f}\int\at{\prt{L}{E}\rund{t}}_{\e=0}\de t+
   2 g\rund{\f} c_{\rm o}\,G\rund{\f}
   \int\at{\prt{L}{W}\rund{t}}_{\e=0}\de t\ \ .
\edm

As Eq.\ (\ref{eqa200}) must hold for an arbitrary choice of 
$h\rund{\f}$, Eq.\ (\ref{eqa50}) is finally reduced to
\bdm
   q\rund{\f}\equiv 0\ \ \  
   \mbox{on}\ \f\in\eck{\f_{\rm in},\f_{\rm out}}\ \ .
\edm
The partial derivatives of $L$ that enter $q$ are
\beao
   \at{\prt{L}{\g}\rund{\f}}_{\e=0} & = & \sqrt{N}\cdot
   \frac{c_{\rm a}\rund{\f} G\rund{\f}}{\sqrt{W-E^2}}\ \ ,\\
   \at{\prt{L}{E}\rund{\f}}_{\e=0} & = & \sqrt{N}\cdot
   c_{\rm a}\rund{\f} G\rund{\f}
   \eck{E\cdot
   \frac{g\rund{\f}-E}{\sqrt{\rund{W-E^2}^3}}
   -\frac{1}{\sqrt{W-E^2}}}\ \ ,\\
   \at{\prt{L}{W}\rund{\f}}_{\e=0} & = & -\frac{1}{2}\,\sqrt{N}\cdot 
   c_{\rm a}\rund{\f} G\rund{\f}\cdot
   \frac{g\rund{\f}-E}{\sqrt{\rund{W-E^2}^3}}\ \ ,
\eeao
so after dividing by $\sqrt{N/\rund{W-E^2}}$ we have
\beao
   \lefteqn{c_{\rm a}\rund{\f} G\rund{\f} +
   c_{\rm o}\,G\rund{\f} E\cdot
   \frac{\int g\rund{t} c_{\rm a}\rund{t} G\rund{t} \de t-E}
   {W-E^2}-}\\
    & - & c_{\rm o}\,G\rund{\f} - c_{\rm o}\,G\rund{\f} g\rund{\f}
   \frac{\int g\rund{t} c_{\rm a}\rund{t} G\rund{t} \de t-E}
   {W-E^2} \equiv 0\ \ .
\eeao

We first note that, if $G\rund{\f'}=0$ at a position $\f'$, this
condition is fulfilled for an arbitrary finite value of $g\rund{\f'}$,
whereas for $G\rund{\f}\ne 0$ a division by $c_{\rm o}\,G\rund{\f}$
yields
\beq\label{eqa300}
   \frac{c_{\rm a}\rund{\f}}{c_{\rm o}}+E\cdot
   \frac{\int g\rund{t} c_{\rm a}\rund{t} G\rund{t} \de t-E}
   {W-E^2} - 1 - g\rund{\f}
   \frac{\int g\rund{t} c_{\rm a}\rund{t} G\rund{t} \de t-E}
   {W-E^2} \equiv 0\ \ .
\eeq
Furthermore, it is easy to see that relation~(\ref{eqa300}) is
invariant under linear transformations 
\bdm
   g\rund{\f}\rightarrow 
   \wt{g}\rund{\f}:=\alpha\,g\rund{\f}+\beta\ \ ,\ \ \ 
   \alpha\neq 0\ \ ,
\edm
because
\beao
   E & \rightarrow & \wt{E}:=\int\wt{g}\rund{\f} c_{\rm o}\,
   G\rund{\f} \de\f = \alpha\,E+\beta\ \ ,\\
   W & \rightarrow & \wt{W}:=\int\eck{\wt{g}\rund{\f}}^2 c_{\rm o}\,
   G\rund{\f} \de\f = \alpha^2 W+2 \alpha\beta E+\beta^2\ \  
\eeao
and
\bdm
   \wt{E}^2 = \alpha^2 E^2 +2 \alpha\beta E+\beta^2\ \ ,
\edm
\bdm
   \wt{W}-\wt{E}^2 = \alpha^2\rund{W-E^2}\ \ ,
\edm
\bdm
   \int\wt{g}\rund{t} c_{\rm a}\rund{t} G\rund{t} \de t
   -\wt{E} = \alpha\eck{\int g\rund{t} c_{\rm a}\rund{t} 
   G\rund{t} \de t - E}\ \ .
\edm
As a consequence, we can chose $\wt{g}\rund{\f}$ such that
\bdm
   \frac{\int\wt{g}\rund{t} c_{\rm a}\rund{t} G\rund{t} \de t
   -\wt{E}}{\wt{W}-\wt{E}^2}=1\ \ \wedge\ \ \wt{E}=1\ \ ,
\edm
thereby simplifying (\ref{eqa300}) to
\bdm
   \wt{g}\rund{\f}=c_{\rm a}\rund{\f}/c_{\rm o}\ \ .
\edm
From this we derive the stationary points of $Q_1$ with respect to the
weight function $g\rund{\f}$ to be given by
\beq\label{eqa400}
   g\rund{\f} = a c_{\rm a}\rund{\f}+b\ \ ,
\eeq
with arbitrary constants $a\neq 0$ and $b$.

The next step of our calculation is to show that \underline{the} 
stationary point given by relation~(\ref{eqa400}) specifies a 
\underline{local} \underline{maximum} of $\abs{Q_1}$.
We consider an arbitrary deviation $\e\cdot h\rund{\f}$ of
$g\rund{\f}$ from (\ref{eqa400}),
\beq\label{eqa450}
   \g\rund{\f,\e}:=a\cdot\eck{c_{\rm a}\rund{\f}+
   \e\cdot h\rund{\f}}+b\ \ ,
\eeq
and analyse its effect on $Q_1$:
\bdm
   Q_1\rund{\e}:=\frac{\int \g\rund{\f,\e} c_{\rm a}\rund{\f} 
   G\rund{\f} \de\f-\int \g\rund{\f,\e} c_{\rm o}\,
   G\rund{\f} \de\f}{\sqrt{\int\eck{\g\rund{\f,\e}-
   \int \g\rund{t,\e} c_{\rm o}\,G\rund{t} \de t}^2 
   c_{\rm o}\,G\rund{\f} \de\f}}\ \ .
\edm

The denominator on the right hand side of this expression is equal to
zero at some point $\e=\e'$ only if
\bdm
   \g\rund{\f,\e'}=\int \g\rund{t,\e'} c_{\rm o}\,G\rund{t} \de t\ \ ,
\edm
which is equivalent to $\g(\f,\e')$ being a constant in $\f$. That,
in turn, requires $h\rund{\f}=-(1/\e') c_{\rm
a}\rund{\f}+\mbox{const}$, resulting in
\beq\label{eqa500}
   \g\rund{\f,\e}=\alpha\rund{\e} c_{\rm a}\rund{\f}+
   \beta\rund{\e}\ \ ,
\eeq
where $\alpha\rund{\e}$ and $\beta\rund{\e}$ are constants in
$\f$. Equation (\ref{eqa500}), however, implies
$\abs{Q_1\rund{\e}}=\mbox{const}$, as can easily be derived from the
definition of $Q_1\rund{\e}$. In the subsequent investigation we will
therefore, without loss of generality, exclude the case of a vanishing
denominator of $Q_1\rund{\e}$.

Introducing new abbreviations
\beao
   X & := & \int\eck{h\rund{\f}-\int h\rund{t}
   c_{\rm o}\,G\rund{t} \de t}^2 G\rund{\f} \de\f = \\
     & =  & \int \eck{h\rund{\f}}^2 G\rund{\f} \de\f-
   c_{\rm o}\eck{\int h\rund{t} G\rund{t} \de t}^2\ge 0\ \ ,\\
   Y & := & \int h\rund{\f} c_{\rm a}\rund{\f} G\rund{\f} \de\f-
   \int h\rund{\f} c_{\rm o}\,G\rund{\f} \de\f=\\
     & =  & \int\eck{h\rund{\f}-
   \int h\rund{t} c_{\rm o}\,G\rund{t} \de t}
   \eckk{c_{\rm a}\rund{\f}-c_{\rm o}}
   G\rund{\f} \de\f\ \ ,\\
   Z & := & \int\eck{c_{\rm a}\rund{\f}}^2 G\rund{\f} \de\f-
   \int c_{\rm a}\rund{\f} c_{\rm o}\,G\rund{\f} \de\f =\\
     & =  & \int\eck{c_{\rm a}\rund{\f}}^2 G\rund{\f} \de\f-
   c_{\rm o} = \int\eck{c_{\rm a}\rund{\f}-c_{\rm o}}^2
   G\rund{\f}\de\f\ge 0\ \ ,
\eeao
and
\beq\label{eqa600}
   q_1\rund{\e}:=\frac{Z+\e Y}
   {\sqrt{c_{\rm o} Z+2\e c_{\rm o} Y+\e^2 c_{\rm o}X}}
\eeq
we write $Q_1\rund{\e}$ in the form
\bdm
   Q_1\rund{\e} = \frac{a}{\abs{a}}\,q_1\rund{\e}\ \ .
\edm
The first derivative of $q_1\rund{\e}$ in $\e$ is 
\bdm
   \tot{}{\e}\,q_1\rund{\e}=\e\cdot\frac{c_{\rm o} Y^2-c_{\rm o}X Z} 
   {\rund{c_{\rm o} Z-2\e c_{\rm o} Y+\e^2 c_{\rm o}X}^{3/2}}\ \ ,
\edm
where, because of the Schwarz inequality\footnote{
   Given three functions $u,v:\bbbr\rightarrow\bbbr$, 
   $G:\bbbr\rightarrow\bbbr_0^+$ the Schwarz inequality states
   \bdm\textstyle
      \eck{\int u\rund{x} v\rund{x} G\rund{x} \de x}^2
      \le\int\eck{u\rund{x}}^2 G\rund{x} \de x
      \int\eck{v\rund{x}}^2 G\rund{x} \de x\ \ . 
   \edm
   {\em Proof\/}: With the definitions
   \beao
      A&:=&\textstyle\int\eck{u\rund{x}}^2 G\rund{x}\de x\ge 0\ \ ,\\
      B&:=&\textstyle\int u\rund{x} v\rund{x} G\rund{x}\de x\ \ ,\\
      C&:=&\textstyle\int\eck{v\rund{x}}^2 G\rund{x}\de x \ge 0\ \ ,
   \eeao
   and an arbitrary $\alpha\in\bbbr$ we have
   \bdm\textstyle
      0 \le C\int\!\eck{u\rund{x}+\alpha v\rund{x}}^2 G\rund{x}\de x = 
      \rund{B+\alpha C}^2\!+A C-B^2\ ,
   \edm
   so that by choosing $\alpha=-B/C$ we obtain $A\cdot C-B^2\ge 0$ or
   equivalently $B^2 \le A\cdot C$ for $C\ne 0$. For the 
   remaining case of $C=0$ we can write 
   \bdm\textstyle
      0 \le \int\!\eck{u\rund{x}+\alpha v\rund{x}}^2 G\rund{x}\de x = 
      A+2\alpha B\ ,
   \edm
   which must hold for any value of
   $\alpha$. From that we derive $B=0$, which is consistent with 
   $B^2 \le A\cdot C=0$, q.e.d.}, 
the numerator on the right hand side is always lower than or equal to
zero. This means
$q_1\rund{\e}$ is monotonously increasing for $\e<0$ but monotonously
decreasing for $\e>0$. Accordingly, $q_1\rund{\e}$ takes its
\underline{global} maximum at $\e=0$ which is, as $q_1\rund{\e=0}\ge 0$
and $\abs{Q_1\rund{\e}}=\abs{q_1\rund{\e}}$, at least a \underline{local}
maximum of $\abs{Q_1\rund{\e}}$.

At this point we know condition (\ref{eqa400}) to correspond to a
local maximum of $\abs{Q_1}$. The final goal of the following
arguments is to demonstrate that this local maximum is also the
global maximum. Taking into account that 
$\abs{Q_1\rund{\e}}=\abs{q_1\rund{\e}}$ and that the global maximum of
$q_1\rund{\e}$ is located at $\e=0$ with $q_1\rund{0}\ge 0$ it is
sufficient to prove $\vert q_1^*\vert\le q_1\rund{0}$ for
the absolute minimum $q_1^*$ of $q_1\rund{\e}$. Because of
the monotony of $q_1\rund{\e}$ we obtain
\bdm
   q_1^*={\rm min}\left\{\lim_{\e\to -\infty}q_1\rund{\e},
   \lim_{\e\to\infty}q_1\rund{\e}\right\}\ \ ,
\edm
with
\bdm
   \lim_{\e\to\infty}q_1\rund{\e}=-\lim_{\e\to -\infty}q_1\rund{\e}=
   Y/\sqrt{X}\ \ ,
\edm
as can be seen from definition (\ref{eqa600}). Depending on the sign
of $Y$ one of the limits is greater than or equal to zero but lower
than or equal to $q_1\rund{0}$, because $q_1\rund{0}$ is the  absolute
maximum of $q_1\rund{\e}$. Therefore, it is
\bdm
  q_1\rund{0}\ge\abs{\lim_{\e\to\infty}q_1\rund{\e}}=
  \abs{\lim_{\e\to -\infty}q_1\rund{\e}}=\abs{q_1^*}\ \ ,   
\edm
which means $\abs{Q_1}$ is \underline{globally} \underline{maximized}
if the weight function $g\rund{\f}$ of the correlation coefficient
(\ref{eqa10}) is in agreement with Eq.\ (\ref{eqa400}).
\subsection*{A.2\ \ The mean error level}
In analogy to Eq. (\ref{eq140}) one can define the mean error level 
$\ave{P_{\rm o}\rund{r_g\ge R}}$ of the correlation coefficient
$r_g$. A good distinction between the two galaxy distributions
described by $p_{\rm o}\rund{\f}$ and $p_{\rm a}\rund{\f}$ is possible
if the mean error level is low. Therefore, we would
like the weight function $g\rund{\f}$ to minimize 
$\ave{P_{\rm o}\rund{r_g\ge R}}$. For this it is a necessary condition
that the mean error level is stationary with respect to small
variations $\e\cdot h\rund{\f}$ of $g\rund{\f}$:
\beq\label{eqa610}
   \at{\tot{}{\e}\ave{P_{\rm o}\rund{r_{\g}\ge R}}}_{\e=0}=0\ \ ,
\eeq
with $\g\rund{\f,\e}:=g\rund{\f}+\e\,h\rund{\f}$. If we write 
$p_{\rm a}\rund{\f}$ in the form
\bdm
   p_{\rm a}\rund{\f}:=p_{\rm o}\rund{\f}\rund{1+\eta\,\n\rund{\f}}
\edm
and fix $\n\rund{\f}$, then in general $g\rund{\f}$ will depend on 
$\eta$. But as (\ref{eqa610}) must hold for any value of $\eta$, we
have
\beq\label{eqa620}
   \at{\prt{}{\eta}\prt{}{\e}\ave{P_{\rm o}\rund{r_{\g}\ge R}}}
   _{\e=0}=0\ \ .
\eeq

What we want to prove here is that for $\eta\to 0$ both 
Eq.\ (\ref{eqa610}) and  Eq.\ (\ref{eqa620}) hold if $g\rund{\f}$ is
in agreement with Eq.\ (\ref{eqa400}). 
For convenience, let us define the symbols
\beao
   P_{{\rm o}g}\rund{R} & := & 1-P_{\rm o}\rund{r_g\ge R}=
   P_{\rm o}\rund{r_g<R}=\\
    & = & \int\!\!\cdots\!\!\int p_{\rm o}\rund{\f_1}\cdots 
   p_{\rm o}\rund{\f_N}\,
   \theta\rund{R-\frac{1}{N}\sum_{i=1}^N g\rund{\f_i}}\,\de^N \f\ \ ,\\
   P_{{\rm a}g}\rund{R} & := & P_{\rm a}\rund{r_g<R}=
   \int\!\!\cdots\!\!\int p_{\rm a}\rund{\f_1}\cdots 
   p_{\rm a}\rund{\f_N}\, 
   \theta\rund{R-\frac{1}{N}\sum_{i=1}^N g\rund{\f_i}}\,\de^N \f\ \ ,\\
   p_{{\rm o}g}\rund{r} & := & \tot{}{r}P_{{\rm o}g}\rund{r}=
   \int\!\!\cdots\!\!\int p_{\rm o}\rund{\f_1}\cdots 
   p_{\rm o}\rund{\f_N}\,
   \delta\rund{r-\frac{1}{N}\sum_{i=1}^N g\rund{\f_i}}\,\de^N \f\ \ ,\\
   p_{{\rm a}g}\rund{r} & := & \tot{}{r}P_{{\rm a}g}\rund{r}=
   \int\!\!\cdots\!\!\int p_{\rm a}\rund{\f_1}\cdots 
   p_{\rm a}\rund{\f_N}\, 
   \delta\rund{r-\frac{1}{N}\sum_{i=1}^N g\rund{\f_i}}\,\de^N \f\ \ , 
\eeao
with Dirac's delta function $\delta\rund{r}$ and Heaviside's step
function $\theta\rund{r}$. All the integrations have to be performed
over $\eck{\f_{\rm in},\f_{\rm out}}$. A partial integration of the
mean error level yields 
\bdm
   \ave{P_{{\rm o}\g}\rund{r_{\g}\ge r}} = 
   1-\int_{-\infty}^{\infty} 
   P_{{\rm o}\g}\rund{r}p_{{\rm a}\g}\rund{r}\,\de r=
   \int_{-\infty}^{\infty}
   P_{{\rm a}\g}\rund{r}p_{{\rm o}\g}\rund{r}\,\de r\ \ . 
\edm

It is evident that for $\eta\to 0$ the mean error level reaches the
constant value $\ave{P_{{\rm o}\g}\rund{r_{\g}\ge r}}= 
\int P_{{\rm o}g}\rund{r}p_{{\rm o}g}\rund{r}\,\de r=1/2$ for an
arbitrary weight function $g$, so obviously Eq.\ (\ref{eqa610}) is 
fulfilled. Furthermore, we find
\beao
   \lefteqn{
   \at{\prt{}{\eta}\prt{}{\e}
   \ave{P_{{\rm o}\g}\rund{r_{\g}\ge r}}}_{\e=0\atop\eta=0}=
   \at{\int\prt{}{\eta}\eck{p_{{\rm o}\g}\rund{r}
   \prt{}{\e}P_{{\rm a}\g}\rund{r}+P_{{\rm a}\g}\rund{r}
   \prt{}{\e}p_{{\rm o}\g}\rund{r}}\de r}_{\e=0\atop\eta=0}=}\\
    & = & \at{\int\prt{}{\eta}\eck{p_{{\rm o}\g}\rund{r}
   \prt{}{\e}P_{{\rm a}\g}\rund{r}-p_{{\rm a}\g}\rund{r}
   \prt{}{\e}P_{{\rm o}\g}\rund{r}}\de r}_{\e=0\atop\eta=0}=\\  
    & = &\frac{1}{N}\int\!\!\cdots\!\!\int p_{\rm o}\rund{\f_1}
   \cdots p_{\rm o}\rund{\f_N}\,p_{\rm o}\rund{\f_1'}\cdots
   p_{\rm o}\rund{\f_N'}
   \,\delta\rund{\frac{1}{N}\sum_{i=1}^N g\rund{\f_i}-
   \frac{1}{N}\sum_{j=1}^N g\rund{\f_j'}}\times\\
    & &\times\sum_{k=1}^N h\rund{\f_k}\sum_{l=1}^N 
   \n\rund{\f_l}\,\de^N \f\,\de^N \f'-
   \frac{1}{N}\int\!\!\cdots\!\!\int p_{\rm o}\rund{\f_1}
   \cdots p_{\rm o}\rund{\f_N}\,p_{\rm o}\rund{\f_1'}\cdots
   p_{\rm o}\rund{\f_N'}\times\\
    & &\times\,\delta\rund{\frac{1}{N}\sum_{i=1}^N g\rund{\f_i}-
   \frac{1}{N}\sum_{j=1}^N g\rund{\f_j'}}
   \sum_{k=1}^N h\rund{\f_k}\sum_{l=1}^N 
   \n\rund{\f_l'}\,\de^N \f\,\de^N \f'=\\
    & = &\eta\int\!\!\cdots\!\!\int p_{\rm o}\rund{\f_1}\cdots
   p_{\rm o}\rund{\f_N}\,p_{\rm o}\rund{\f_1'}\cdots
   p_{\rm o}\rund{\f_N'}\times\\
    & &\times\,\delta\rund{\frac{1}{N}\sum_{i=1}^N 
   \ECK{g\rund{\f_i}-g\rund{\f_i'}}}
   \sum_{j=1}^N h\rund{\f_j}
   \sum_{k=1}^N \ECK{\n\rund{\f_k}-\n\rund{\f_k'}}\,
   \de^N \f\,\de^N \f'\ \ . 
\eeao

Because of the delta function only those points
$(\f_1,\ldots,\f_N,\f_1',\ldots,\f_N')$ contribute to the integral
which meet the condition 
\beq\label{eqa1000}
   \sum_{i=1}^N \ECK{g\rund{\f_i}-g\rund{\f_i'}}=0\ \ .
\eeq
Now suppose $g\rund{\f}$ to be in accordance with expression
(\ref{eqa400}). Then (\ref{eqa1000}) implies
\bdm
   \sum_{k=1}^N \ECK{\n\rund{\f_k}-\n\rund{\f_k'}}=0\ \ ,
\edm
and it is
\bdm
   \at{\prt{}{\eta}\prt{}{\e}\ave{P_{{\rm o}\g}
   \rund{r_{\g}\ge r}}}_{\e=0\atop\eta=0}=0\ \ .
\edm

This final result shows that the weight function specified by
relation (\ref{eqa400}) not only maximizes the quantity $\abs{Q_1}$ as
discussed in the preceding section, but is also close to a stationary
point of the mean error level, if $\eta\approx 0$, i.e. 
$p_{\rm a}\rund{\f}\approx p_{\rm o}\rund{\f}$.  
%
%
%
%

%
%
%
%

\begin{thebibliography}{}
%
\bibitem[1991]{msb91}
   Bartelmann~M., Schneider~P., 1991, A\&A 248, 349 
%
\bibitem[1992]{msb92}
   Bartelmann~M., Schneider~P., 1992, A\&A 259, 413 
%
\bibitem[1993a]{msb93a}
   Bartelmann~M., Schneider~P., 1993a, A\&A 268, 1 
%
\bibitem[1993b]{msb93b}
   Bartelmann~M., Schneider~P., 1993b, A\&A 271, 421 
%
\bibitem[1994]{msb94}
   Bartelmann~M., Schneider~P., 1994, A\&A 284, 1 (BS)
%
\bibitem[1995]{msb95}
   Bartelmann~M., 1995, A\&A 298, 661 
%
\bibitem[1995]{for95}
   Fort~B., Mellier~Y., Dantel-Fort~M., Bonnet~H., Kneib~J.-P., 1995,
   A\&A in press 
%
\bibitem[1988]{fug88}
   Fugmann~W., 1988, A\&A 204, 73
%
\bibitem[1990]{fug90}
   Fugmann~W., 1990, A\&A 240, 11
%
\bibitem[1995]{hut95}
   Hutchings~J.~B., 1995, Astron.\ J.\ 109, 928
%
\bibitem[1981]{kuehr}
   K\"uhr~H., Witzel~A., Pauliny-Toth~I.~I., Nauber~U., 1981,
   A\&AS 45, 367 
%
\bibitem[1993]{pad}
   Padmanabhan~T., ``Structure formation in the universe'', 
   Cambridge University Press, 1993
%
\bibitem[1994]{rod94}
   Rodrigues-Williams~L.~L., Hogan~C.~J., 1994, Astron.\ J.\ 107, 451
%
\bibitem[1995]{sei95}
   Seitz~S., Schneider~P., 1995, A\&A 302, 9
%
\bibitem[1993]{sti93}
   Stickel~M., K\"uhr~H., Fried~J.~W., 1993, A\&AS 97, 483
%
\bibitem[1993a]{sti93a}
   Stickel~M., K\"uhr~H., 1993a, A\&AS 100, 395
%
\bibitem[1993b]{sti93b}
   Stickel~M., K\"uhr~H., 1993b, A\&AS 101, 521
%
\bibitem[1995]{wu95}
   Wu~X.-P., Han~J., 1995, M.N.R.A.S. 272, 705
%
\end{thebibliography}
\end{document}